\documentclass[twocolumn, pra, superscriptaddress]{revtex4-1}
\usepackage{amssymb}
\usepackage{amsmath}
\usepackage{epsfig}
\usepackage{graphics, graphicx}
\usepackage{bbold}
\usepackage{psfrag}
\usepackage{mathcomp}
\usepackage{subfigure}
\usepackage{verbatim}
\usepackage{color}
\usepackage[colorlinks,citecolor=blue]{hyperref}
\def\cp#1{\mathbf{#1}}

\begin{document}

\title{Nature of polaron-molecule transition in Fermi polarons}
\author{Cheng Peng}
\thanks{These authors contributed equally to this work.}
\affiliation{Beijing National Laboratory for Condensed Matter Physics, Institute of Physics, Chinese Academy of Sciences, Beijing 100190, China}
\affiliation{School of Physical Sciences, University of Chinese Academy of Sciences, Beijing 100049, China}
\author{Ruijin Liu}
\thanks{These authors contributed equally to this work.}
\affiliation{Beijing National Laboratory for Condensed Matter Physics, Institute of Physics, Chinese Academy of Sciences, Beijing 100190, China}
\author{Wei Zhang}
\affiliation{Department of Physics, Renmin University of China, Beijing 100872, China}
\affiliation{Beijing Key Laboratory of Opto-electronic Functional Materials and Micro-nano Devices, Renmin University of China,
Beijing 100872, China}
\author{Xiaoling Cui}
\email{xlcui@iphy.ac.cn}
\affiliation{Beijing National Laboratory for Condensed Matter Physics, Institute of Physics, Chinese Academy of Sciences, Beijing 100190, China}
\affiliation{Songshan Lake Materials Laboratory, Dongguan, Guangdong 523808, China}
\date{\today}

\begin{abstract}
It has been commonly believed that a polaron to molecule transition occurs in three-dimensional(3D) and two-dimensional(2D) Fermi polaron systems as the attraction between the single impurity and majority fermions gets stronger. The conclusion has been drawn from the separate treatment of polaron and molecule states and thus deserves a close reexamination. In this work, we explore the polaron and molecule physics by utilizing a unified variational ansatz with up to two particle-hole(p-h) excitations(V-2ph). 
We confirm the existence of a first-order transition in 3D and 2D Fermi polarons, and show that the nature of such transition lies in an energy competition between systems with different momenta ${\cp Q}=0$ and $|{\cp Q}|=k_F$,  here ${\cp Q}$ is defined as the momentum of Fermi polaron system with respect to the Fermi sea of majority fermions (with Fermi momentum $k_F$). 
The literally proposed molecule ansatz is identified as an asymptotic limit of $|{\cp Q }|=k_F$ state in strong coupling regime, which implies a huge $SO(3)$(for 3D) or $SO(2)$ (for 2D) ground state degeneracy in this regime. The recognization of such degeneracy is crucially important for evaluating the molecule occupation in realistic systems with finite impurity density and at finite temperature. To compare with recent experiment of 3D Fermi polarons, we have calculated various physical quantities under the V-2ph framework and obtained results that are in good agreements with experimental data in the weak coupling and near resonance regime.  Further, to check the validity of our conclusion in 2D, we have adopted a different variational method based on the Gaussian sample of 
high-order p-h excitations(V-Gph), and found the same conclusion on the nature of polaron-molecule transition therein.
For 1D system, the V-2ph method predicts no sharp transition and the ground state is always at ${\cp Q}=0$ sector, consistent with exact Bethe ansatz solution. The presence/absence of polaron-molecule transition is analyzed to be closely related to the interplay effect of Pauli-blocking and p-h excitations in different dimensions.
\end{abstract}
\maketitle

\section{Introduction}

Polaron refers to a typical quasi-particle in highly polarized systems. Its concept was first raised by Landau in the 1930s, when he discussed how an electron moving through a solid will cause the distortion of the lattice and get trapped\cite{Landau}. After nearly a century, the concept of polaron has been well acknowledged and extended to various physical systems. In particular, in recent years the ultracold atoms have served as an ideal platform for the study of polaron physics,  thanks to the high controllability of species, number and interaction therein. One important branch of these studies is the Fermi polaron, which describes an impurity immersed in and dressed by a fermonic environment. To date, the attractive and repulsive Fermi polarons have been extensively explored in ultracold atoms both experimentally\cite{Zwierlein,Salomon,Salomon2,Grimm,Kohl,Grimm2016,Roati,Sagi} and theoretically\cite{Chevy, Lobo, Combescot1, Combescot2, Prokofev, Leyronas, Punk,ChevyM4, Enss, Bruun, Castin, Parish, Parish2, Edwards, Parish4, ZhangWei, Pethick, ZhangWei2, MC_2d_1,MC_2d_2,MC_2d_3,Combescot3,
Cui, Troyer, Bruun2, Parish3, Demler}. 

For the attractive Fermi polaron in high dimensions, it has been commonly believed that a polaron to molecule transition occurs when the attraction between the impurity and majority fermions increases\cite{Prokofev, Leyronas, Punk, Enss, Bruun, Castin, Parish,Parish2, MC_2d_1, MC_2d_2}. Namely, depending on the attraction  strength between the impurity and fermions,  it could end up with two distinct destinies: one destiny is that the impurity is dressed with the surrounding cloud of majority fermions and forms a fermionic {\it polaron}; the other is that the impurity essentially binds with one single fermion on top of the Fermi surface to form a bosonic {\it molecule}.  To characterize these distinct pictures, the following variational ansatz for polaron and molecule states with truncated $n$ particle-hole(p-h) excitations have been proposed\cite{Chevy, ChevyM4, Combescot1, Combescot2, Leyronas, Punk, Castin, Parish, Parish2, Cui, Parish3, Pethick, Parish4}:
\begin{equation}
P_{2n+1}(0)=\left[\psi_0 c^{\dag}_{{\cp 0} \downarrow}+\sum_{l=1}^n \sum_{{\cp k}_i {\cp q}_j}\psi_{{\cp k}_i {\cp q}_j} c^{\dag}_{{\cp P} \downarrow}\prod_{i=1}^l c^{\dag}_{{\cp k}_i\uparrow}\prod_{j=1}^l c_{{\cp q}_j \uparrow}\right] |{\rm FS}\rangle_{N}; \label{wf_p}
\end{equation}
%\begin{equation}
%P_{2n+1}({\cp Q})=\left[\psi_0 c^{\dag}_{{\cp Q} \downarrow}+\sum_{l=1}^n \sum_{{\cp k}_i {\cp q}_j}\psi_{{\cp k}_i {\cp q}_j} c^{\dag}_{{\cp P} \downarrow}\prod_{i=1}^l c^{\dag}_{{\cp k}_i\uparrow}\prod_{j=1}^l c_{{\cp q}_j \uparrow}\right] |{\rm FS}\rangle_{N}; \label{wf_p}
%\rho(\Delta E) \sim k_F^2/\sqrt{\Delta E}
%\end{equation}
\begin{eqnarray}
M_{2n+2}(0)&=&\left[\sum_{{\cp k}}\phi_{{\cp k}} c^{\dag}_{-{\cp k},\downarrow}c^{\dag}_{\cp k,\uparrow}+  \right. \nonumber\\
&&\left. \sum_{l=1}^{n} \sum_{{\cp k}_i {\cp q}_j}\phi_{{\cp k}_i {\cp q}_j} c^{\dag}_{{\cp P} \downarrow}\prod_{i=1}^{l+1} c^{\dag}_{{\cp k}_i\uparrow}\prod_{i=1}^l c_{{\cp q}_j \uparrow}\right] |{\rm FS}\rangle_{N-1}.\nonumber\\
\label{wf_m}.
\end{eqnarray}
Here $c^{\dag}_{\bf k,\sigma}$ is the creation operator of spin-$\sigma$ fermions at momentum ${\bf k}$ and the $\downarrow$-spin is the impurity, $|\rm FS\rangle_{\textit{N}}$ is the Fermi sea of $\uparrow$-spin with number $N$; all ${\bf q}$ (${\bf k}$) are below (above) the Fermi surface of $\uparrow$-atoms and ${\cp P}=\sum_j{\cp q}_j-\sum_i{\cp k}_i$. The two ansatz above  have been shown to lead to a first-order transition between polaron and molecule for both 3D\cite{Leyronas, Punk, Enss, Bruun, Castin} and 2D\cite{Parish,Parish2} Fermi polaron systems. The same conclusion was also drawn from Monte-Carlo methods\cite{Prokofev, MC_2d_1,MC_2d_2}, where the polaron and molecule were treated separately with their energies extracted from different physical quantities. %based on the separate treatment of polaron and molecule states
%been widely accepted for both 3D studied\cite{..} and .   accepted by the community , based on an energy crossing from the two ansatz above\cite{..} or from other quantities as extracted in Monte-Carlo methods\cite{..}.

%The existence of a first-order transition between $|P(0)\rangle$(polaron) and $|M(0)\rangle$(molecule) has been widely accepted by the community, based on an energy crossing from the two ansatz above\cite{..} or from other quantities as extracted in Monte-Carlo methods\cite{..}.

The separate treatment of polaron and molecule, though physically inspiring, has its own drawback as the transition appears to be artificially designed at the very beginning. As a result, the conclusion of polaron-molecule transition can easily get questioned. For instance, a previous theory\cite{Edwards} claimed the absence of such transition by showing that the two types of variational ansatz are mutually contained in a generalized momentum space if more p-h excitations are included. Therefore, the relation and competition between polaron and molecule deserve a close re-examination under a unified framework.

On the experimental side, the polaron-molecule transition has been identified by a continuous zero-crossing of quasi-particle residue, instead of a sudden jump as in the first-order transition, in both 3D and 2D Fermi gases\cite{Zwierlein,Kohl}. In particular,  a recent experiment on 3D Fermi polarons has observed a smooth evolution of various physical quantities across the  polaron-molecule transition, as well as a coexistence of polaron and molecule near their transition\cite{Sagi}. All these observations need to be reconsidered carefully following the unified treatment of polaron and molecule states.

With above motivations, in a recent work\cite{Cui2} we have adopted a unified variational method with one p-h excitations(V-1ph) to study the Fermi polaron problem in 3D. Specifically, the unified ansatz we used is $P_3({\bf Q})$, i.e., the extension of $P_3(0)$ in Eq.~(\ref{wf_p}) to finite momentum. Note that here momentum ${\cp Q}$ is defined in the reference frame of background Fermi sea of all majority atoms; in other words, it represents the momentum difference between the ground states of interacting system (Fermi polaron) and non-interacting system (zero-momentum impurity plus the majority Fermi sea), and thus can well characterize the interaction effect.
By this, we found that the bare molecule state $M_2(0)$ actually constitutes part of $P_3({\bf Q})$ with $|{\bf Q}|=k_F$ (denoted as $P_3(k_F)$ for short), here $k_F$ the Fermi momentum of majority fermions. Due to the incomplete variational space of  $M_2(0)$ even within the lowest-order p-h excitations, it always has a higher energy than $P_3(k_F)$. The significance of introducing $M_2(0)$ is found to lie in the strong coupling regime, where it can serve as a good approximation for $P_3(k_F)$. Within V-1ph method, we concluded that the nature of ``polaron-molecule transition" is given by an energy competition between $P_3(0)$ and $P_3(k_F)$. %, while at the transition point $M(0)$ can well approximate the serve as a good approximation for $P(k_F)$.
This naturally resolves the theoretical debate in Ref.\cite{Edwards} because the transition is between different $Q$-states rather than between different forms of variational ansatz.
Furthermore, near the transition point, we found the double-minima (at $|{\bf Q}|=0$ and $k_F$) structure of the impurity dispersion curve, providing the underlying mechanism for polaron-molecule coexistence in realistic systems. Based on this, we qualitatively explained the smooth polaron-molecule transition as observed in the recent experiment\cite{Sagi} with a finite impurity density and at finite temperature. % continuous change of residue and a smoot this regime in practical system.  and thus a smoothened transition in realistic cold atoms experiment.

%***********************************
\begin{figure}[t]
\includegraphics[width=0.5\textwidth]{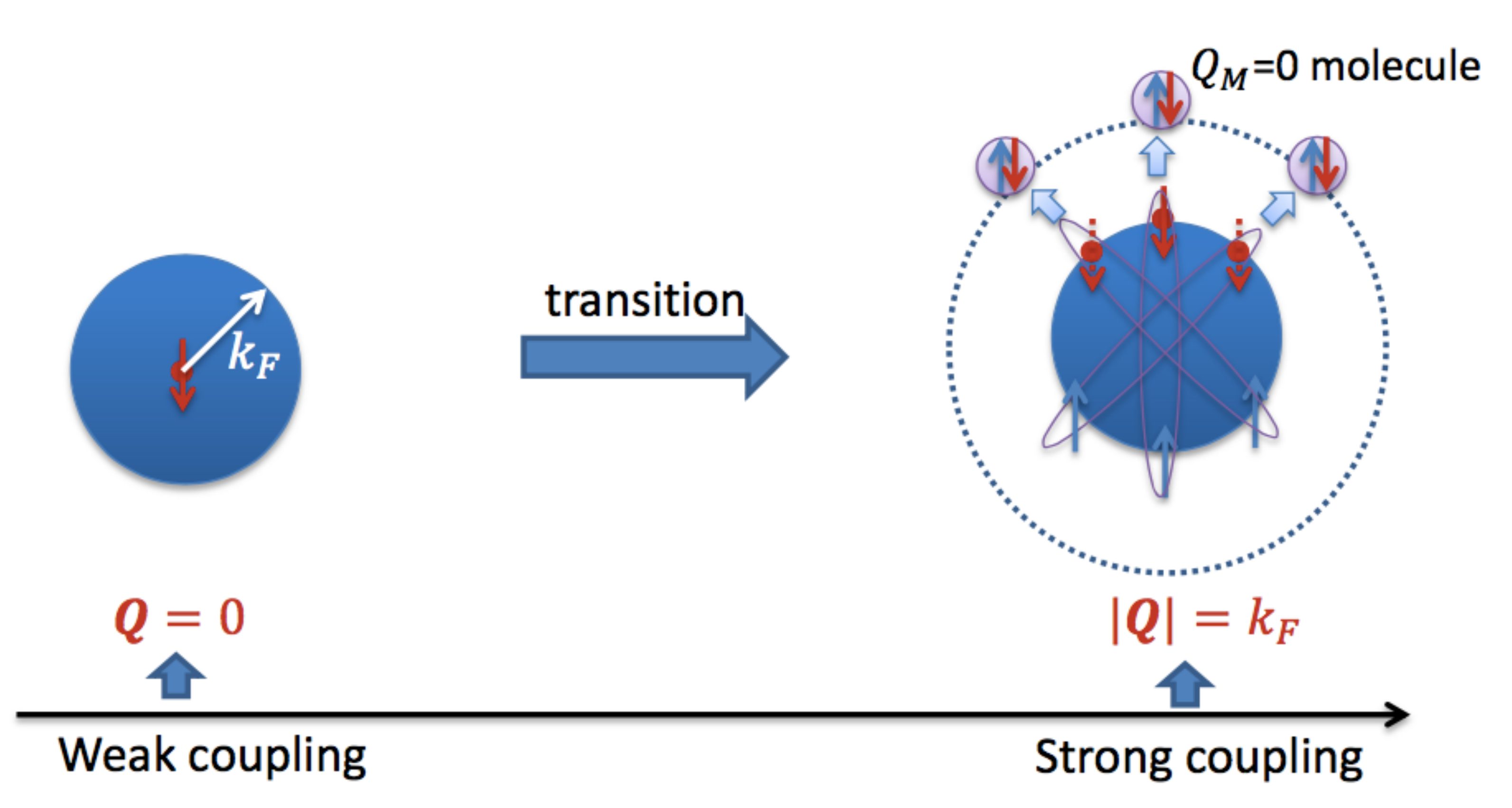}
\caption{(Color online). Illustration of polaron-molecule transition as changing the impurity($\downarrow$)-fermion($\uparrow$) attraction from weak (left side) to strong (right side). In the weak coupling regime, the ground state is a zero-momentum polaron (${\cp Q}=0$) dominated by a zero-momentum impurity dressed by particle-hole excitations in the background Fermi sea. In the strong coupling regime, the ground state switches to $|{\cp Q}|=k_F$, in order to facilitate the impurity pairing with a fermion originally at the Fermi surface to form a deeply bound molecule with zero center-of-mass momentum ($Q_M=0$).  This results in a huge ground state degeneracy in the molecule regime, i.e., SO(3) for 3D and SO(2) for 2D. In this sense, the molecule ansatz $M(0)$  represents a symmetry-breaking state within the degenerate manifold.  }\label{fig_schematic}
\end{figure}
%***********************************

In the present work, we extend the study of Fermi polaron problem to various dimensions using the unified variational method with up to two p-h excitations (V-2ph), namely, under variational ansatz $P_5({\cp Q})$. %, which are expected to produce more accurate results on the competition between polaron and molecule.
%Previously, the zero-momentum case (under $P_5(0)$) has been studied in Ref.\cite{Combescot2,Parish2}, and here we extend to arbitrary ${\bf Q}$.
With V-2ph method, we confirm the existence of polaron-molecule transition in 3D and 2D, and re-enforce the conclusion made in Ref.\cite{Cui2} that the nature of such transition lies in an energy competition between different momenta ${\cp Q}=0$ and $|{\cp Q}|=k_F$. %, i.e, between
%   , as revealed initially in Ref.\cite{..} with lowest-order ansatz, still preserves when adding two p-h excitations. Namely, the transition is due to an energy competition between
%$P_5(0)$ and $P_5(k_F)$.
%while at the transition point $P_5(k_F)$ can be well approximated by the molecule state $M_4(0)$. %The significance of introducing molecule lies in the strong coupling regime where it can serve as a good approximation for the $Q=k_F$ state.
%Importantly, the identification of molecule state as holding total momentum $Q=k_F$   means that the ground state in strong coupling side is highly degenerate due to the rotational invariance of ${\bf Q}$, which resemblance the case with an isotropic spin-orbit coupling\cite{..}.
%Same as Ref.\cite{Cui2}, near the transitions the two momenta states appear as double minima in the dispersion curve, signifying the coexistence of two states under realistic situation. 
Here, we find the main effect of including two p-h excitations is to shift the transition point and the coexistence region to weaker coupling regime, from which we obtain a reasonably better prediction to various physical quantities as measured in the weak coupling and resonance regime of Fermi polaron experiment\cite{Sagi}.
Moreover, we emphasize in this work an important fact that has been overlooked by previous studies, i.e., the molecule ground state has a huge degeneracy ($SO(3)$ for 3D and $SO(2)$ for 2D). %, as the vector ${\bf k}_F$ can point to any direction.    that in the molecule regime the ground state of single-impurity system has a huge $SU(2)$(for 3D) or $SO(2)$(for 2D) degeneracy, as the vector ${\bf k}_F$ can point to any direction. Physically, this large degeneracy is because the impurity can select any particle at the Fermi surface to form a molecule with zero center-of-mass momentum. We will show that t
The recognization of such degeneracy is crucially important for correctly evaluating the individual occupation of polaron and molecule in their coexistence region for realistic Fermi polaron systems. In Fig.\ref{fig_schematic}, we illustrate the nature of polaron-molecule transition as well as the origin of huge ground state degeneracy for molecules.

%Moreover, we emphasize that the identification of ground state in the molecule side as holding finite momentum $Q=k_F$ implies it is highly degenerate (due to the rotational invariance of total momentum), which crucially affect the "boson" occupation in the realistic situation with finite impurity density. Finally, we show how the polaron-molecule transition can be smoothened by a finite impurity density and a finite temperature, and we also propose to use the momentum-resolved radio-frequency spectroscopy to probe the polaron-molecule transition and coexistence in realistic cold atoms experiments.

To further check the validity of our results in 2D, we adopt a different variational method based on the Gaussian sample of high order p-h excitations(V-Gph)\cite{Shi}, which gives the same conclusion for the nature of polaron-molecule transition therein. For 1D system, the V-2ph method predicts no sharp transition and the ground state is always the $Q=0$ state for any coupling strength, consistent with the Bethe ansatz solutions. These comparisons further justify the validity of V-2ph method and the reliability of our results in various dimensions. We analyze that the presence or absence of polaron-molecule transition is closely related  to the interplay effect of Pauli-blocking and p-h excitations in different dimensions.

The rest of the paper is organized as follows. In Sec.II, we present the algorithm from two variational ansatz to treat the Fermi polaron problem: one is the variational ansatz with up to two p-h excitations(V-2ph), and the other is the Gaussian variational ansatz with high order p-h excitations(V-Gph). In Sec.III, we present the results of polaron-molecule transition for single impurity system in various dimensions from the two methods, and analyze the intrinsic reason for the presence/absence of such transition in different dimensions. In Sec. IV, we use the single-impurity results to investigate the coexistence and smooth crossover between polaron and molecule in 3D Fermi polaron systems, in comparison with the experimental data from Ref.\cite{Sagi}.  %we  reveal the smooth conversion between polaron and molecule in realistic cold atoms experiment using the momentum-resolved radio-frequency spectroscopy.
Finally the results are summarized in Sec. V.

\section{Methods}

We consider the following Hamiltonian describing a spin-$\downarrow$ impurity interacting with spin-$\uparrow$ majority fermions:
\begin{equation}
\label{eq:H}
H=\sum_{\mathbf{k}\sigma} \epsilon_{\mathbf{k},\sigma} c_{\mathbf{k} \sigma}^{\dagger} c_{\mathbf{k} \sigma}+g/L^d \sum_{\mathbf{Q}, \mathbf{k}, \mathbf{k}^{\prime}} c_{\mathbf{Q}-\mathbf{k}, \uparrow}^{\dagger} c_{\mathbf{k}, \downarrow}^{\dagger} c_{\mathbf{k}^{\prime}, \downarrow} c_{\mathbf{Q}-\mathbf{k}^{\prime}, \uparrow}
\end{equation}
where $\epsilon_{\mathbf{k}}=\mathbf{k}^2/(2m)$; $d$ is the dimension of the system; $g$ is the bare coupling constant which needs to be renormalized in 2D and 3D due to the induced ultraviolet divergence in two-body scattering process. Specifically, for 3D $g$ is related to the s-wave scattering length $a_s$ via $1/g=m/(4\pi a_s)-1/V \sum_{\mathbf{k}}1/(2\epsilon_{\mathbf{k}})$ with $V=L^3$ the volume of the system; for 2D, the scattering length $a_{2d}$ defines the two-body binding energy $E_{2b}=-1/ma_{2d}^2$ and $g$ is related to $E_{2b}$ via $1/g=-1/S \sum_{\mathbf{k}}1/(2\epsilon_{\mathbf{k}}-E_{2b})$ where $S=L^2$ is the area of the system. In this work we  take $\hbar$ as unity for brevity. %In the following, we take $V=1$ and $S=1$ unless otherwise stated.

In this section, we present the algorithm of two variational methods used to treat Fermi polaron problems. One is the the unified variational ansatz $P_5({\bf Q})$ with up to two p-h excitations(V-2ph), in comparison with the molecule ansatz $M_4({\bf Q}_M)$. The other is the Gaussian variational ansatz with high order p-h excitations(V-Gph).

\subsection{Unified variational approach with up to two p-h excitations (V-2ph)}
%It is straightforward to extend the zero-momentum ansatz for polaron and molecule, as presented in Eqs.(\ref{wf_p},\ref{wf_p}), to arbitrary finite momentum.
In the following, we will present the algorithm of $P_5({\bf Q})$, the polaron ansatz with arbitrary momentum and with up to two p-h excitations, as well as %its resulted integrated equation in various dimensions. Then we present
the algorithm of $M_4({\bf Q}_M)$, the molecule ansatz with arbitrary momentum and with one p-h excitations.
It is noted that the ${\cp Q}=0$ case of $P_5({\cp Q})$ have been studied previously in 3D\cite{Combescot2,Leyronas}, 2D\cite{Parish2},  and 1D\cite{Combescot3} Fermi polaron systems;  the ${\cp Q}_M=0$ case of $M_4({\cp Q}_M)$ have also been studied previously in 3D\cite{Leyronas, Punk,ChevyM4} and 2D\cite{Parish,Parish2} systems. Here we generalize the study to arbitrarily finite momenta, which evolves more numerical work than the zero-momentum case. % heavier numerical work than It is straightforward to generalize the derivation of the equations therein to arbitrary ${\cp Q}$ and ${\cp Q}_M$.
The intrinsic relation between the two ansatz will also be discussed.

\subsubsection{$P_5({\bf Q})$}
The generalized polaron ansatz with up to two p-h excitations is:
\begin{eqnarray}
P_{5}({\bf Q})&=&\left[\psi_0 c^{\dag}_{{\bf Q} \downarrow}+\sum_{{\cp k} {\cp q}}\psi_{{\cp k}  {\cp q} } c^{\dag}_{{\cp Q}+{\cp q}-{\cp k} \downarrow} c^{\dag}_{{\cp k}\uparrow}c_{{\cp q} \uparrow} + \right. \nonumber\\
&&\left.  \frac{1}{4}\sum_{{\cp k}{\cp k'} {\cp q}{\cp q'}}\psi_{{\cp k}{\cp k'}  {\cp q}{\cp q'} } c^{\dag}_{{\cp Q}+{\cp q}+{\cp q'}-{\cp k}-{\cp k'} \downarrow} c^{\dag}_{{\cp k}\uparrow}c^{\dag}_{{\cp k'}\uparrow}c_{{\cp q} \uparrow}c_{{\cp q'} \uparrow}\right] |{\rm FS}\rangle_{N}. \nonumber\\ \label{wf_p_Q}
\end{eqnarray}
By imposing the Schr{\" o}dinger equation, we can obtain the coupled integral equations for all variational coefficients, from which the ground state energy can be obtained. This is equivalent to minimizing the energy functional $E_{\rm tot}=\langle H \rangle $ for a normalized ansatz. In this paper, we take the unperturbed Fermi sea $|\rm FS\rangle_{\textit{N}}$ as the reference system and define the impurity energy as $E=E_{\rm tot}-E_{\rm FS}$, with $E_{\rm FS}$ is the energy of $|\rm FS\rangle_{\textit{N}}$.

\begin{widetext}
The equations for the variational coefficients are:
%\begin{widetext}
\begin{eqnarray}
\label{eq0}
-\frac{1}{g}(E-E^{(0)}_{\mathbf{Q}})\psi _0&=&\sum_{\mathbf{kq}}\psi _{\mathbf{kq}};\\
%\label{eq1}
-\frac{1}{g}(E-E^{(1)}_{\mathbf{kq}})\psi _{\mathbf{kq}}&=&\psi _0+ \sum_{\mathbf{K}} \psi _{\mathbf{Kq}}-
\sum_{\mathbf{q}'} \psi _{\mathbf{kq'}}-\sum_{\mathbf{Kq'}}\psi _{\mathbf{kKqq'}};\\
%\label{eq2}
-\frac{1}{g}(E-E^{(2)}_{\mathbf{kk}'\mathbf{qq}'})\psi _{\mathbf{kk}'\mathbf{qq}'}&=&-\psi _{\mathbf{kq}}-\psi _{\mathbf{k}'\mathbf{q}'}+\psi _{\mathbf{k}\mathbf{q}'}+\psi _{\mathbf{k}'\mathbf{q}}+ \sum_{K} \psi _{\mathbf{Kk}'\mathbf{qq}'}+\sum_{\mathbf{K}} \psi _{\mathbf{kKqq}'}
-\sum_{\mathbf{Q'}} \psi _{\mathbf{kk}'\mathbf{Q'q}'}-\sum_{\mathbf{Q'}} \psi _{\mathbf{kk}'\mathbf{qQ'}},
\end{eqnarray}
where $E_{\cp Q}^{(0)}=\epsilon_{\mathbf{Q}}$,  $E^{(1)}_{\mathbf{kq}}=\epsilon_{\mathbf{Q+q-k}}
+\epsilon_{\mathbf{k}}-\epsilon_{\mathbf{q}}$,\ $E^{(2)}_{ \mathbf{k}\mathbf{k}'\mathbf{q}\mathbf{q}'} = \epsilon_{\mathbf{Q}+\mathbf{q}+\mathbf{q}' -\mathbf{k}- \mathbf{k}'} + \epsilon_{\mathbf{k}} +\epsilon_{\mathbf{k}'}  - \epsilon_{\mathbf{q}} - \epsilon_{\mathbf{q}'}$.
%\end{widetext}
As before, all ${\bf q}$ (${\bf k}$) in these equations are by default below (above) the Fermi surface of $|\rm FS\rangle_{\textit{N}}$.
\end{widetext}

Above equations can be solved in 1D using iterative method. For 2D and 3D, due to the renormalization scheme of bare coupling $g$, the equations can be simplified by using

\begin{eqnarray}
&&g\sum_{k'} \frac{\alpha_{{\cp k'}{\cp q}}}{E-E^{(2)}_{\mathbf{k}\mathbf{k}'\mathbf{q}\mathbf{q}'}}\sim 0;\nonumber\\
&&g\sum_{k'} \frac{1}{E-E^{(2)}_{\mathbf{k}\mathbf{k}'\mathbf{q}\mathbf{q}'}}\sim 1.\nonumber
\end{eqnarray}
The final equations for numerical simulation are
\begin{eqnarray}
E &=& \epsilon_{\cp Q} + \sum_{\mathbf{q}} A_{\cp q};  \label{eq1}\\
{\alpha}_{\mathbf{k} \mathbf{q} } &=& \frac{ {A}_{\mathbf{q}} -\sum_{\mathbf{q}'} {G}(\mathbf{k},\mathbf{q},\mathbf{q}') }{E-E^{(1)}_{\mathbf{k} \mathbf{q}}} ; \label{eq2}\\
{G}(\mathbf{k},\mathbf{q},\mathbf{q}') &=& \frac{ {\alpha}_{\mathbf{k} \mathbf{q}'} - {\alpha}_{\mathbf{k} \mathbf{q} }  - \sum_{\mathbf{k}'} \frac{ {G}(\mathbf{k}',\mathbf{q},\mathbf{q}') }{ E-E^{(2)}_{\mathbf{k}\mathbf{k}'\mathbf{q}\mathbf{q}'} } }{h(\mathbf{k},\mathbf{q},\mathbf{q}')}, \label{eq3}
\end{eqnarray}

with
\begin{eqnarray}
{A}_{\mathbf{q}} &=& \frac{ 1- \sum_{\mathbf{k} \mathbf{q}'} \frac{{G}(\mathbf{k},\mathbf{q},\mathbf{q}')}{E-E^{(1)}_{\mathbf{k} \mathbf{q}}}  }{h(\mathbf{q})};\\
h(\mathbf{q}) &=& \frac{1}{g} - \sum_{\mathbf{k}} \frac{1}{E-E^{(1)}_{\mathbf{k} \mathbf{q}}} %= \frac{m}{4\pi a_s} - \sum_{\mathbf{k}} \frac{1}{2\epsilon_{\mathbf{k}} }- \sum_{\mathbf{k}} \frac{1}{E-E^{(1)}_{\mathbf{k} \mathbf{q}}}
;  \\
h(\mathbf{k},\mathbf{q},\mathbf{q}') &=& \frac{1}{g} - \sum_{\mathbf{k}'} \frac{1}{E-E^{(2)}_{\mathbf{k}\mathbf{k}'\mathbf{q}\mathbf{q}'}}, %=  \frac{m}{4\pi a_s} - \sum_{\mathbf{k}} \frac{1}{2\epsilon_{\mathbf{k}} } - \sum_{\mathbf{k}} \frac{1}{E-E^{(2)}_{\mathbf{k}\mathbf{k}'\mathbf{q}\mathbf{q}'}}
\end{eqnarray}
where we have defined  ${\alpha}_{\mathbf{k}\mathbf{q}} = \psi_{\mathbf{k}\mathbf{q}}/ \psi_0$,  ${A}_{\mathbf{q}} = g (1+\sum_{\cp k}{\alpha}_{\mathbf{k}\mathbf{q}})$, ${G}(\mathbf{k},\mathbf{q},\mathbf{q}') = g\sum_{\cp k'}\psi_{\mathbf{k}\mathbf{k'}\mathbf{q}\mathbf{q}'}/\psi_0$.

Due to the rotational invariance of momentum ${\cp Q}$, in this work we have taken it along $z$ axis for simplicity. Compared to the zero momentum case, here the finite ${\cp Q}$ in 3D and 2D introduces more momentum variables in the simulation and thus requires a heavier numerical work. In practice, we have used iterative scheme to solve Eqs.~(\ref{eq1}) to (\ref{eq3}). In updating $E$ in Eq.~(\ref{eq1}) and updating $G({\cp k},{\cp q},{\cp q'})$ in Eq.~(\ref{eq3}), we have used the  successive over-relaxation method to reduce the fluctuation and ensure the convergency of the results.

\subsubsection{$M_4({\bf Q}_M)$}

The generalized molecule ansatz with one p-h excitations is written as:
\begin{eqnarray}
M_{4}({\cp Q}_{\rm M})&=&\left[\sum_{{\cp k}}\phi_{{\cp k}} c^{\dag}_{{\cp Q_{\rm M}}-{\cp k},\downarrow}c^{\dag}_{\cp k,\uparrow}+  \right. \nonumber\\
&&\left.  \frac{1}{2} \sum_{{\cp k}{\cp k'} {\cp q}}\phi_{{\cp k}{\cp k'}  {\cp q} } c^{\dag}_{{\cp Q}_{\rm M}+{\cp q}-{\cp k}-{\cp k'} \downarrow} c^{\dag}_{{\cp k}\uparrow}c^{\dag}_{{\cp k'}\uparrow}c_{{\cp q} \uparrow}\right] |{\rm FS}\rangle_{N-1}. \nonumber\\ \label{wf_m}
\end{eqnarray}
By imposing the Schr{\" o}dinger equation, one can obtain the equations for all variables $\phi_{{\cp k}}$, $\phi_{{\cp k}{\cp k'}  {\cp q} }$. Again for 2D and 3D cases, the equations can be simplified. Namely, by introducing two auxiliary functions $\gamma=g\sum_{\cp k} \phi_{\cp k}$ and $\eta_{{\cp k}{\cp q}}=g\sum_{\cp k'}\phi_{{\cp k}{\cp k'}{\cp q}}$, we can arrive at the following integral equations for $\tilde{\eta}_{{\cp k}{\cp q}}=\eta_{{\cp k}{\cp q}}/\gamma$ (see the  $Q_{\rm M}=0$ case in \cite{Leyronas,Parish,Punk,Parish2}):
\begin{eqnarray}
&&\frac{1}{g}-\sum_{\cp k}\frac{1}{E+E_F-E^{(1)}_{{\cp k}}}=\sum_{{\cp k}{\cp q}}  \frac{ \tilde{\eta}_{\mathbf{k}\mathbf{q}} }{ E+E_F-E^{(1)}_{\mathbf{k}}} ;\\
&&\left[\frac{1}{g}-\sum_{\cp k'}\frac{1}{E+E_F-E^{(2)}_{{\cp k}{\cp k'}{\cp q}}}\right]\tilde{\eta}_{\mathbf{k}\mathbf{q}} = -\frac{1+\sum_{\cp q'}\tilde{\eta}_{\mathbf{k}\mathbf{q'}} }{E-E^{(1)}_{\cp k}}- \nonumber\\
&& \ \ \ \ \ \ \ \ \ \ \ \ \ \ \ \ \ \ \ \ \ \ \ \ \ \ \ \ \ \sum_{\mathbf{k}'} \frac{ \tilde{\eta}_{\mathbf{k}\mathbf{q}} }{ E+E_F-E^{(2)}_{\mathbf{k}\mathbf{k}',\mathbf{q}}},
\end{eqnarray}
with $E^{(1)}_{\cp k}=\epsilon_{{\cp Q}_M-{\cp k}}+\epsilon_{{\cp k}}$ and $E^{(2)}_{\mathbf{k}\mathbf{k}',\mathbf{q}}=\epsilon_{{\cp Q}_M-{\cp k}-{\cp k'}+{\cp q}}+\epsilon_{{\cp k}}+\epsilon_{{\cp k}'}-\epsilon_{{\cp q}}$.

Again in the calculation we take ${\cp Q}_M$ along $z$ axis due to its rotational invariance. Compared to $P_5({\cp Q})$, the simulation of $M_4({\cp Q}_M)$ is easier due to the smaller variational space. One can obtain the molecule energy $E$ either by using iterative method or by solving large matrix equations with respect to $\tilde{\eta}_{\mathbf{k}\mathbf{q}}$. We have confirmed that these two methods produce consistent results.

\subsubsection{Relation between $P_5({\bf Q})$ and $M_4({\bf Q}_M)$}

In our previous work\cite{Cui2}, we have discussed the intimate relation between $M_2(0)$ and $P_3({\cp Q})$ with $|{\cp Q}|=k_F$. The discussion can be % shown that up to the single p-h excitations on top of the unperturbed Fermi sea, the variational space of bare molecule $M_2(0)$ is incomplete and only constitutes part of $P_3(k_F)$, i.e., the polaron ansatz  at finite momentum $k_F$. In fact, the relation of $M(0)$ and $P(k_F)$ is generic and can be
straightforwardly extended to other momentum sectors and to arbitrary levels of p-h excitations. Here we consider the case of $P_5({\bf Q})$ and $M_4({\bf Q}_M)$ and discuss their relation as below. We start with the following equality between two Fermi sea states%$|FS\rangle_{N-1}=c_{{\cp k_F}\uparrow}|FS\rangle_N$,
\begin{equation}
|\rm FS\rangle_{\textit{N}-1}=c_{{\cp k}_F \uparrow}|\rm FS\rangle_\textit{N}. \label{relation}
\end{equation}
%where ${\cp e}_{Q}$ is unit vector of direction ${\cp Q}$.
Here ${\cp k}_F$ is the Fermi momentum that can point to any direction on the Fermi surface. Given (\ref{relation}), one can see that if we further take
\begin{equation}
\psi_0=0, \ \psi_{{\cp k}  {\cp q} }=\phi_{\cp k}\delta_{{\cp q},{\cp k}_F},\ \psi_{{\cp k}{\cp k'}  {\cp q}{\cp q'}}=\phi_{{\cp k}{\cp k'}  {\cp q} }\delta_{{\cp q'},{\cp k}_F}, \label{corr1}
\end{equation}
then $P_{5}({\bf Q})$ in (\ref{wf_p_Q}) exactly reproduces $M_{4}({\cp Q}_{\rm M})$ in (\ref{wf_m}) under the relation
\begin{equation}
{\cp Q}_{\rm M}={\cp Q}+{\cp k}_F.  \label{corr2}
\end{equation}
Eqs.~(\ref{corr1},\ref{corr2}), which can be directly generalized to arbitrary order of p-h excitations, immediately tell us two important facts:

(i) $M_4({\cp Q}_{\rm M})$ has a smaller variational space than $P_5({\cp Q}={\cp Q}_{\rm M}-{\cp k}_F)$. Specifically, the former corresponds to only considering a particular configuration of p-h excitations in the latter, i.e., with one hole pinning at the Fermi surface [see Eq.~(\ref{corr1})]. In principle, such configuration is not isolated and can be coupled to other p-h excitations via interactions, which will further reduce the variational energy. Due to such incomplete variational space, $M_4({\cp Q}_{\rm M})$ always has a higher variational energy than $P_5({\cp Q}={\cp Q}_{\rm M}-{\cp k}_F)$ for the ground state of the system. When reduced to the special case ${\cp Q}_{\rm M}=0$ and $|{\cp Q}|=k_F$, we arrive at the conclusion that $M_4(0)$ always produces a higher energy than $P_5({\cp Q})$ with $|{\cp Q}|=k_F$. This is a direct extension of the conclusion in our previous work with one p-h excitations\cite{Cui2}.

(ii) The correspondence (\ref{corr2}) tells that, the previously studied {\it zero-momentum} molecule  $M(0)$ actually stays in a different momentum sector from the {\it zero-momentum} polaron $P(0)$. Such momentum difference, ${\cp k}_F$, which originates from the relation (\ref{relation}) between two Fermi seas $|{\rm FS}\rangle_N$ and $|{\rm FS}\rangle_{N-1}$, is robust against the choice of reference state. Nevertheless, to correctly characterize the status of the impurity, it is important to choose the reference state as $|{\rm FS}\rangle_N$, instead of $|{\rm FS}\rangle_{N-1}$. By choosing $|{\rm FS}\rangle_N$ as the reference state, the momenta of $P(0)$ and $M(0)$ are respectively ${\cp Q}=0$ and ${\cp Q}=-{\cp k}_F$, giving the momentum difference ${\cp k}_F$.
%it is For instance, in this work we have chosen the reference state as the ground state of non-interacting system, $|\Psi\rangle_0=c^{\dag}_{0,\downarrow}|{\rm FS}\rangle_N$, and then the momenta of $P(0)$ and $M(0)$ are respectively $0$ and $-{\cp k}_F$, giving the momentum difference ${\cp k}_F$; alternatively, one can choose  $|{\rm FS}\rangle_{N-1}$ as the reference state, and then the momenta of $P(0)$ and $M(0)$ are respectively ${\cp k}_F$ and $0$, again giving the momentum difference ${\cp k}_F$. 
Because of such momentum difference, $M(0)$ and $P(0)$ should have zero overlap (note that the Hamiltonian (\ref{eq:H}) preserves the total momentum). %This momentum difference is robust against the choice of reference state.
 Recognizing such difference is crucially  important for understanding the nature of polaron-molecule transition, as addressed in section \ref{result}.

Based on (i,ii), we can conclude that up to two p-h excitations, the generalized polaron ansatz $P_5({\cp Q})$ can serve as the unified variational wave function for both polaron and molecule states. The ground state of the system can then be obtained by searching for the energy minimum in the ${\cp Q}$-space.  %actually to search for the global minimum of polaron dispersion over all momenta.
%Surely, one may argue that one can equally use the finite-momentum molecule ansatz as the unified variational ansatz because it can also cover the variational space of polaronic state if include more p-h excitations. This argument was exactly presented in previous study\cite{Cui2} to make a conclusion that the molecule and polaron ansatz are mutually contained with each other and thus there is no transition in this system. This conclusion, however, is incorrect because it missed the momentum difference between the two ansatz and thus misunderstood the nature of such transition. This is related to the second point as illustrated below.

\subsection{Gaussian variational method with high-order particle-hole excitations (V-Gph)}

For 2D system, besides the V-2ph method we adopt the Gaussian variational method with high-order p-h excitations (V-Gph)\cite{Shi}. The essence of this method is the combination of fermionic Gaussian state\cite{Bravyi,Kraus} and the Lee-Low-Pines (LLP) transformation\cite{Lee}. To be self-contained, in the following we give a brief introduction to this method.

Applying the  LLP transformation $U_{\rm LLP}=e^{-i\mathbf{\hat{K}\hat{r}}}$, where $\mathbf{\hat{K}}=\sum_{\mathbf{k}} \mathbf{k}c_{\mathbf{k}\uparrow}^\dag c_{\mathbf{k}\uparrow}$ is the total momentum of the background spin-up atoms and $\mathbf{\hat{r}}$ is the coordinate of the impurity, the  Hamiltonian (\ref{eq:H}) can be transformed as
%$\mathbf{\hat{r}}$ and $\mathbf{\hat{p}}$ denote the coordinate and momentum operator of the impurity, respectively, the
\begin{eqnarray}
H_{\rm LLP}&=&U_{\rm LLP}^\dag H U_{\rm LLP}\nonumber\\
&=&\sum_{\mathbf{k}}(\epsilon_{\mathbf{k}}-\mu)c_{\mathbf{k}
\uparrow}^{\dag}c_{\mathbf{k}\uparrow}+\frac{\mathbf{\hat{p}}^2}{2m}
-\sum_{\mathbf{k}}\frac{\mathbf{\hat{p}\cdot k}}{m}c_{\mathbf{k}
\uparrow}^{\dag}c_{\mathbf{k}\uparrow}\nonumber\\
&&+\sum_{\mathbf{k,k'}}\frac{\mathbf{k\cdot k'}}{2m}c_{\mathbf{k}\uparrow}^{\dag}c_{\mathbf{k}
\uparrow}c_{\mathbf{k'}\uparrow}^{\dag}
c_{\mathbf{k'}\uparrow}
+\frac{g}{L^2}\sum_{\mathbf{k,k'}}
c_{\mathbf{k}\uparrow}^{\dag}c_{\mathbf{k'}\uparrow}.\nonumber\\
\label{eq:HLLP}
\end{eqnarray}
Here $\mathbf{\hat{p}}$ is the momentum operator of the impurity. Note that here we have introduced an additional term ``$-\mu\sum_{\mathbf{k}}c_{\mathbf{k}\uparrow}^{\dag}c_{\mathbf{k}\uparrow}$" into the original Hamiltonian Eq.~(\ref{eq:H}) to tune the particle number of the background Fermi sea. After the LLP transformation, the conserved total momentum of the system transforms into the momentum of the impurity, i.e.,
\begin{eqnarray}
U_{\rm LLP}^\dag(\mathbf{\hat{p}}+\mathbf{\hat{K}})U_{\rm LLP}=\mathbf{\hat{p}}.
\label{eq:momentum_total}
\end{eqnarray}
Thus we can replace $\mathbf{\hat{p}}$ in $H_{\rm LLP}$ with its eigenvalue $\mathbf{Q}$, which eliminates the degree of the impurity.

We further use fermionic Gaussian state to approximate the ground state with total momentum $\mathbf{Q}$ of the transformed Hamiltonian, Eq.~(\ref{eq:HLLP}). The fermionic Gaussian state is defined as
\begin{eqnarray}
|\Psi_{\rm GS}\rangle=c_{\mathbf{Q}\downarrow}^{\dag}U_{\rm GS}|0\rangle,
\label{GS_state}
\end{eqnarray}
where $|0\rangle$ is chosen to be the vacuum state and
\begin{equation}
U_{\rm GS}=e^{i\frac{1}{4}A^{T}\xi A} \label{u}
\end{equation}
is called the Gaussian unitary operator, $A=(a_{1,\mathbf{k}_1},\ldots,a_{1,\mathbf{k}_{N_k}},a_{2,\mathbf{k}_1},\ldots,a_{2,\mathbf{k}_{N_k}})^{T}$,
 $N_k$ is the number of $\mathbf{k}$ modes satisfying $|\mathbf{k}|\leq k_c$ with cutoff $k_c$, the Majorana operators are defined as
$a_{1,\mathbf{k}_j}=c_{\mathbf{k}_j,\uparrow}^{\dag}+c_{\mathbf{k}_j,
\uparrow}$, $a_{2,\mathbf{k}_j}=i(c_{\mathbf{k}_j,\uparrow}^{\dag}-c_{\mathbf{k}_j,\uparrow})$,
and the variational parameter $\xi$ is an antisymmetric Hermitian matrix which has $2N_k^2-2N_k$ free matrix elements. We point out that the use of Majorana operators is just for computational convenience and the operators can be re-expressed in terms of $c_{\mathbf{k}_j,\uparrow}^{\dag}$ and $c_{\mathbf{k}_j,\uparrow}$ as in Ref\cite{Dolgirev}.

To eliminate the gauge degree of freedom in $\xi$, it is convenient to introduce a covariance matrix~\cite{Shi}
\begin{eqnarray}
(\Gamma)_{s_1,\mathbf{k}_1;s_2,\mathbf{k}_2}=\frac{i}{2}\langle\Psi_{\rm GS}|[a_{s_1,\mathbf{k}_1},a_{s_2,\mathbf{k}_2}]|\Psi_{\rm GS}\rangle,
\label{eqn:gammam}
\end{eqnarray}
with $s_1(s_2)=1,2$. The covariance matrix is related to $\xi$ as
\begin{eqnarray}
\Gamma=-U_{m}\left(
  \begin{array}{cc}
    \mathbf{0} & -\mathbf{1}_{N_k} \\
    \mathbf{1}_{N_k} & \mathbf{0} \\
  \end{array}
\right) U_{m}^{T},
\label{eqn:covariant}
\end{eqnarray}
where $U_m=e^{i\xi}$ and $\mathbf{1}_{N_k}$ is the identity matrix of dimension $N_k$.

By reversing the LLP transformation, the eigenstate of the original Hamiltonian (\ref{eq:H}) with a total conserved momentum $\mathbf{Q}$ can be expressed as a non-Gaussian state
\begin{eqnarray}
|\Psi\rangle=U_{\rm LLP}c_{\mathbf{Q}\downarrow}^{\dag}U_{\rm GS}|0\rangle.
\label{NGS_state}
\end{eqnarray}
The imaginary-time evolution equation for the non-Gaussian state Eq.~(\ref{NGS_state}) can be written as
\begin{eqnarray}
d_{\tau}|\Psi \rangle&=&-\mathcal{P}(H-E_{\rm tot})|\Psi\rangle,\label{ima3}
\end{eqnarray}
where $\mathcal{P}$ is the projection operator onto the subspace spanned by tangent vectors of the variational manifold, $E_{\rm tot}=\langle\Psi |H|\Psi \rangle$ can be calculated using Wick's theorem. Finally we obtain
\begin{widetext}
\begin{eqnarray}
E_{\rm tot}&=&\frac{1}{2}\sum_{\mathbf{k}}\varepsilon_{\mathbf{k}}-\frac{\mu N_{\mathbf{k}}}{2}+\frac{1}{4}\sum_{\mathbf{k}}(\varepsilon_{\mathbf{k}}-\mu-\frac{\mathbf{Q\cdot k}}{m})
(\Gamma_
{1,\mathbf{k};2,\mathbf{k}}-\Gamma_{2,\mathbf{k};1,\mathbf{k}})+\frac{\mathbf{Q}^2}{2m}\nonumber\\
&&+\frac{g}{2L^2}N_{\mathbf{k}}+\frac{g}{4L^2}\sum_{\mathbf{k, k'}}(\Gamma_
{1,\mathbf{k};2,\mathbf{k}'}-\Gamma_{2,\mathbf{k};1,\mathbf{k}'})+\frac{1}{8m}\sum_{\mathbf{k}}\mathbf{k}^2\nonumber\\
&&+\frac{1}{32m}[\sum_{\mathbf{k}
}\mathbf{k}(\Gamma_{1,\mathbf{k};2,\mathbf{k}}-\Gamma_{2,\mathbf{k};1,\mathbf{k}})]^2
-\frac{1}{8m}\sum_{\mathbf{k}
,\mathbf{k}'}\mathbf{k\cdot k'}\Gamma_{1,\mathbf{k};1,\mathbf{k}'}
\Gamma_{2,\mathbf{k};2,\mathbf{k}'}+\frac{1}{8m}\sum_{\mathbf{k}
,\mathbf{k}'}\mathbf{k\cdot k'}\Gamma_{1,\mathbf{k};2,\mathbf{k}'}
\Gamma_{2,\mathbf{k};1,\mathbf{k}'}.
\end{eqnarray}
\end{widetext}
To be consistent with the variational approach with truncated p-h excitations, we calculate the energy $E=E_{\rm tot}+\mu N_{\uparrow}-E_{\rm FS}$.
The imaginary time equation of motion (EOM) for the covariance matrix $\Gamma$ is
\begin{equation}
\partial _{\tau }\Gamma=-h-\Gamma h\Gamma,
\label{eqn:EOM}
\end{equation}
with
\begin{eqnarray}
h=4\frac{\delta E_{\rm GS}}{\delta\Gamma}.
\label{hm}
\end{eqnarray}
Evolving $\Gamma$ according to Eq.~(\ref{eqn:EOM}) until the variational energy converges, we can finally obtain the approximated ground state.

Now we discuss the level of p-h excitations in V-Gph. Since the Fermi sea $|\textrm{FS}\rangle_N$ is also a Gaussian state, we can replace $|0\rangle$ as $|\textrm{FS}\rangle_N$ in Eq.~(\ref{GS_state}) and immediately one can see that it can include multiple p-h excitations. By expanding $U_{\rm GS}$ in terms of $\xi$: $U_{\rm GS}=1+i\frac{1}{4}A^{T}\xi A+...$, the wave function $\Psi$ can also be expanded in terms of $\xi$. We note that the first two terms in the expansion have included all the bare and one p-h excitation terms in $P_3({\cp Q})$, while the coefficients of two and higher p-h excitation terms in $\Psi$ are strongly correlated with those of one p-h terms and thus are not free variables. This means that V-Gph can be a better variational approach than V-1ph, but not necessarily better than V-2ph. In this work, we use it as a complementary method to test the reliability of V-2ph.

\section{Polaron-molecule transition/crossover for single impurity systems} \label{result}

In this section, we study the polaron to molecule transition or crossover for single impurity systems  in various dimensions. We will apply the V-2ph method for all dimensions, in combination with  V-Gph method for 2D and the Bethe-ansatz method for 1D.  The conclusion for the presence/absence of polaron-molecule transition from these methods are consistent. % include the identification of the existence/absence and the nature of polaron-molecule transition for the ground state of single-impurity system, as well as the smooth crossover between polaron and molecule physics in realistic systems with finite impurity density and at finite temperature.

\subsection{3D}

In our previous work\cite{Cui2}, we have used the V-1ph method based on ansatz $P_3({\cp Q})$ to unveil the nature of polaron-molecule transition in 3D. Here by using V-2ph method with up to two p-h excitations, we will re-examine the polaron and molecule physics in this system. In our numerical simulations, we have taken the momentum cutoff as $k_c=30k_F$.

%***********************************
\begin{figure}[h]
\centering
\includegraphics[width=0.5\textwidth]{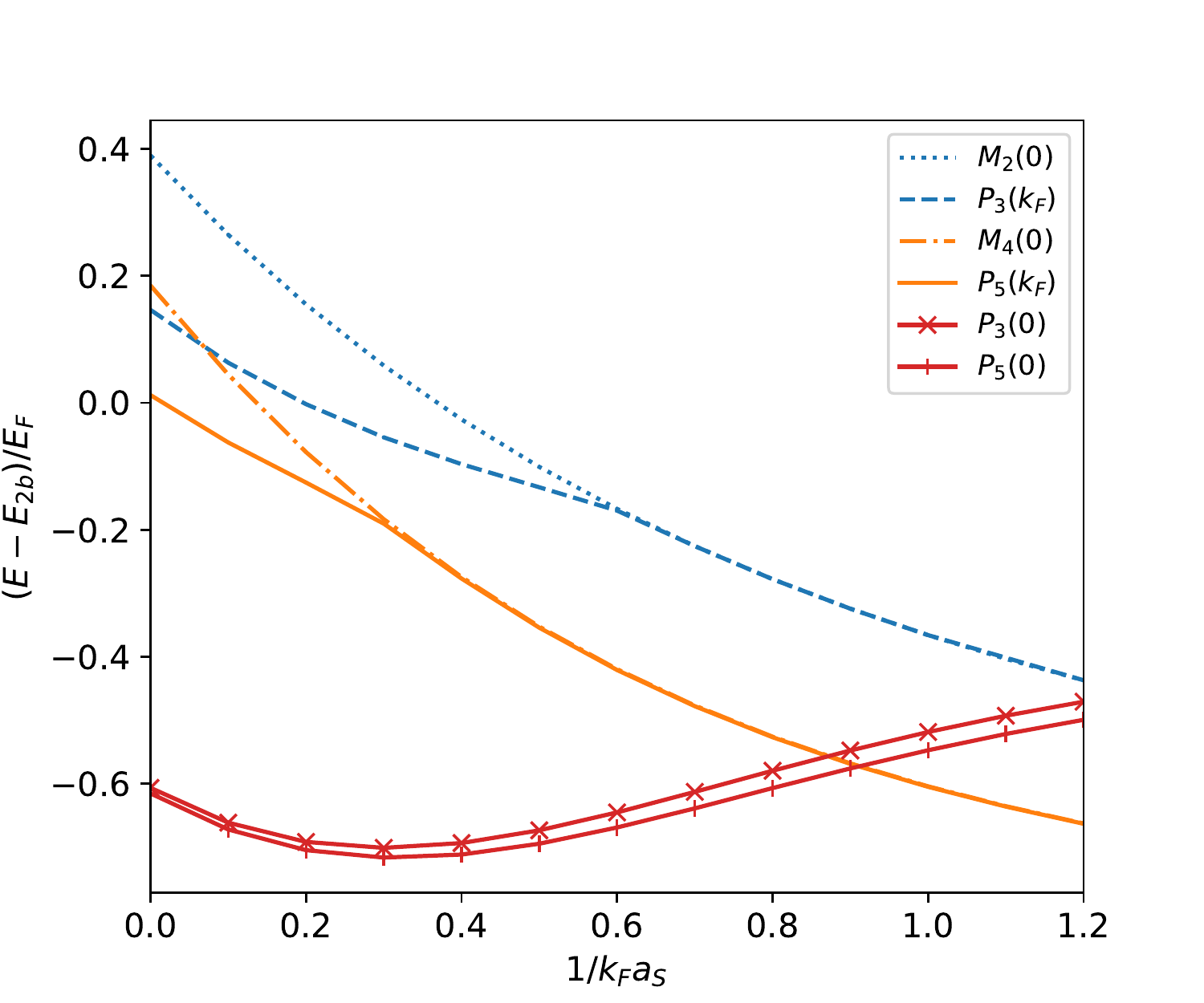}
\caption{(Color online). Energy comparison between various ansatz for 3D single impurity system. %$P_5({\cp k}_F)$, $M_4(0)$, $P_3({\cp k}_F)$, $M_2(0)$, $P_5(0)$ and $P_3(0)$.
All energies are shifted by $E_{2b}=-1/(ma_s^2)$ in $a_s>0$ side in order to highlight the difference.}
\label{fig_mol_3D}
\end{figure}
%***********************************

%***********************************

\begin{figure}[h]
\centering
\includegraphics[width=0.5\textwidth]{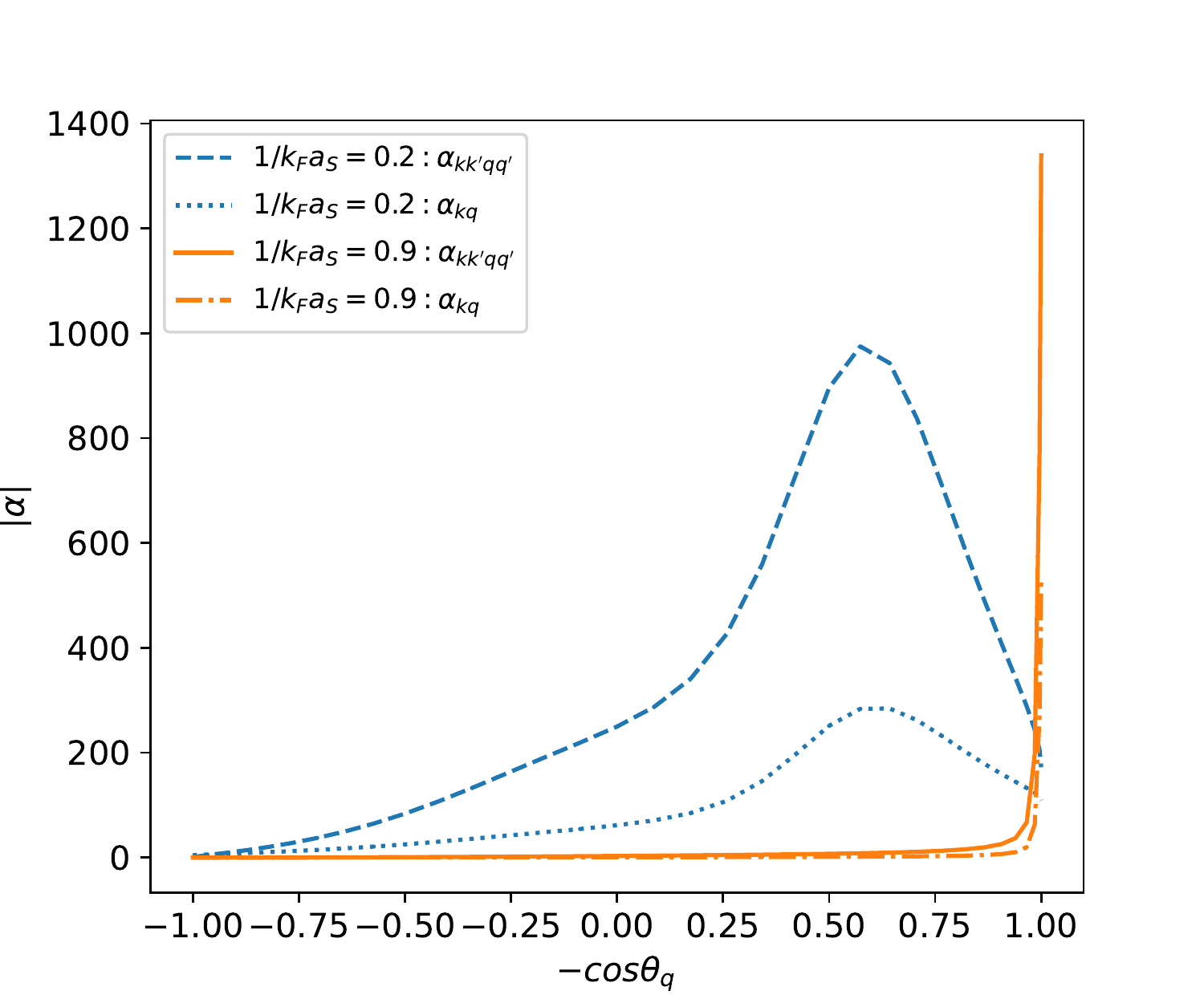}
\caption{(Color online). Hole angular distribution of variational coefficients in $P_5(k_F)$ at different coupling strengths. Here we use the polar coordinate ($|{\cp k}|,\theta_k,\phi_k$) to characterize momentum ${\cp k}$, with $\theta_k\in[0,\pi)$ and $\phi_k\in[0,2\pi)$. In the figure we choose  ${\cp k}=(1.32k_F, 0.53, 0.44),\  {\cp k'} = (2.64k_F, 0.53, 0.44),\  {\cp q'}= (k_F, 0, 0.44)$, and ${\cp q} = (k_F,\theta_q,0.44)$ in $\alpha_{\mathbf{k}\mathbf{q}}\equiv \psi_{\mathbf{k}\mathbf{q}}/\psi_0 $ and $\alpha_{ \mathbf{k}\mathbf{k}'\mathbf{q}  \mathbf{q}' }\equiv \psi_{\mathbf{k}\mathbf{k}'\mathbf{q}  \mathbf{q}' }/\psi_0$.} 
\label{fig_wf}
\end{figure}
%***********************************

First, we investigate the relation between $M_4(0)$ and $P_5({\cp Q})$ with ${\cp Q}=k_F{\cp e}_z$, and we will denote the latter state as $P_5(k_F)$ for short.  As discussed in above section, due to the incomplete variational space of $M_4(0)$, it should be energetically unfavorable as compared to $P_5(k_F)$.  In Fig.\ref{fig_mol_3D}, we show their energies, in comparison with $P_3(k_F)$ and $M_2(0)$, as functions of coupling strength.  It is found that the molecule state $M_4(0)$ (or $M_2(0)$) always has a higher energy than $P_5(k_F)$ (or $P_3(k_F)$), as expected. Only in the strong coupling side, the energy difference between $M_4(0)$ and $P_5(k_F)$ (or between $M_2(0)$ and $P_3(k_F)$) becomes invisible. For instance, $M_4(0)$ energetically approaches $P_5(k_F)$ at couplings $1/(k_Fa_s)\gtrsim 0.3$, and $M_2(0)$ energetically approaches $P_3(k_F)$ at $1/(k_Fa_s)\gtrsim 0.6$. Moreover, we can see that the V-2ph method produces a lower energy for both polaron and molecule states, as compared to those from V-1ph method.

To explain why the energies of $M_4(0)$ and $P_5(k_F)$ become so close in the strong coupling limit, we examine the wave-function of $P_5(k_F)$ in Fig.\ref{fig_wf}. Specifically, we show the hole angular distribution of variational coefficients
%$\alpha_{\bar{\mathbf{k}}\mathbf{q}} $ and $\alpha_{\bar{ \mathbf{k} } \bar{\mathbf{k}}'\mathbf{q}  \bar{\mathbf{q}}' }$ in  $P_5({\cp k}_F)$
at two different coupling strengths. It is found that at intermediate coupling $1/{k_Fa_s}=0.2$, the angular distribution of the hole ($\cp q$) spreads in a broad region,  while at stronger coupling $1/{k_Fa_s}=0.9$ the distribution shows a pronounced peak at $\theta_q=\pi$, i.e., along the opposite direction of ${\cp Q}(=k_F{\cp e}_z)$. Recalling Eqs.~(\ref{corr1},\ref{corr2}), this corresponds to locking the hole at $-{\cp Q}$ so as to produce a molecule state with ${\cp Q}_{\rm M}=0$. We have checked that such pronounced hole distribution at $-{\cp Q}$ applies for general excited momenta ${\cp k}$ and ${\cp k'}$. Together with the energy resemblance as shown in Fig.\ref{fig_mol_3D}, this serves as a strong evidence that $M_4(0)$ indeed can well approximate $P_5(k_F)$ in the strong coupling limit.

%***********************************
\begin{figure}[h]
\centering
\includegraphics[width=0.5\textwidth]{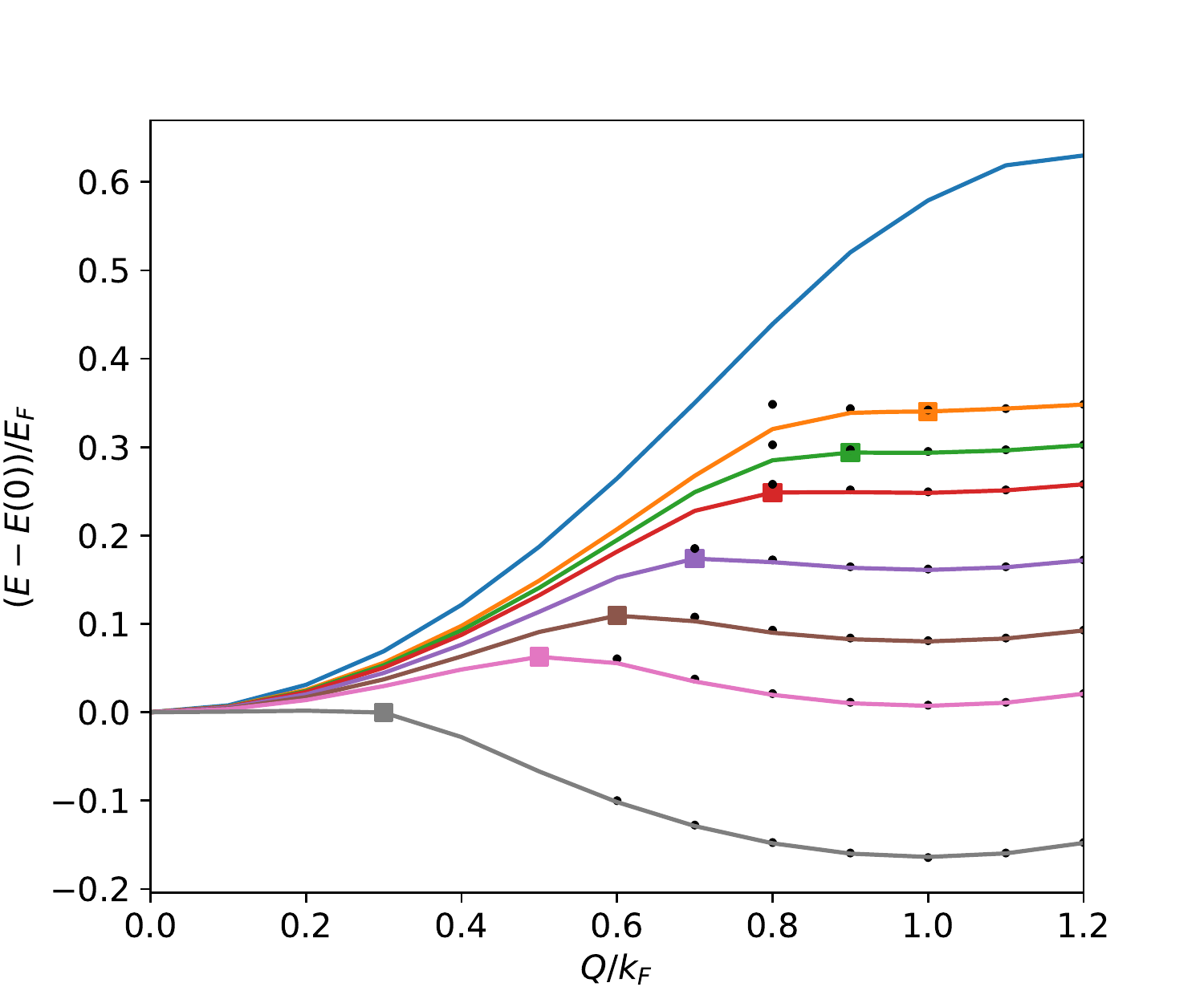}
\caption{(Color online). (a) Energy dispersion of $P_5(\mathbf{Q})$(solid lines) in 3D at various couplings (from top to bottom) $1/(k_Fa_s)=0.2,\ 0.5,\ 0.55,\ 0.6,\ 0.7,\ 0.8,\ 0.9,\ 1.2$, shifted by the value at $\mathbf{Q}=0$. The rectangular point mark the position of maximum energy, and the small black dots show the energies of $M_4({\cp Q}_{\rm M})$, with $|{\cp Q}_{\rm M}|$ shifted by $k_F$ in order to compare with the energies of $P_5(\mathbf{Q})$. Here $Q=|{\cp Q}|$.
}
\label{fig_transition_3D}
\end{figure}
%***********************************

Given the fact that the molecule $M_4(0)$ is nothing but just a good approximation for the finite-momentum state  $P_5(k_F)$, now we are ready to investigate the polaron-molecule competition by examining the energy  dispersion $E(Q)$ from $P_5({\cp Q})$, with $Q=|{\cp Q}|$ (in our numerical calculation, we have taken ${\cp Q}$ along z-direction). In Fig.\ref{fig_transition_3D}, we show $E(Q)$ for various coupling strengths.
We can see that for weak coupling $1/(k_Fa_s)\lesssim 0.5$, there is only one minimum in the dispersion and $Q=0$ polaron is the only ground state. Near ${\cp Q}\sim 0$, one has
\begin{equation}
E(Q)=\epsilon_P+\frac{{\cp Q}^2}{2m_P^{*}},
\end{equation}
with $\epsilon_P=E(0)$ and $m_P^{*}$ respectively the energy and effective mass of polaron state.
As increasing $1/(k_Fa_s)$ to $\sim 0.5$ and beyond, another minimum appears at $Q=k_F$ as a metastable state.  At $1/k_Fa_s=0.91$, the two minima has the same energy, signifying a first-order transition between $Q=0$ and $Q=k_F$ states, or between polaron and molecule states given that $M_4(0)$ can well approximate $P_5(k_F)$ near the transition (see black dots). At even stronger attractions, the local minimum at $Q=0$ is bended downwards and the only stable state is at $Q=k_F$, the molecule state. It is found that near the local minimum $Q\sim k_F$, the dispersion well follows
\begin{equation}
E(Q)=\epsilon_M+\frac{(|{\cp Q}|-k_F)^2}{2m_M^{*}}, \label{E_M}
\end{equation}
with $\epsilon_M=E(k_F)$ and $m_M^{*}$ respectively the energy and effective mass of molecule state.
Here with V-2ph method, the double minima structure of the dispersion appears in the coupling window $1/k_Fa_s\in (0.5,1.2)$,  moving to weaker coupling side as compared to the double minima region from V-1ph method\cite{Cui2}.

In Fig.\ref{fig_mol_3D}, we compare the energies at two momenta $0$ and $k_F$ from both V-2ph and V-1ph methods. One can see that under V-2ph, the critical point for the transition is at $(1/k_Fa_s)_c=0.91$, very close to the critical point obtained from Monte-Carlo\cite{Prokofev} and diagrammatic\cite{Leyronas} methods. Clearly, this critical point shifts to weaker coupling side as compared to the value $(1/k_Fa_s)_c=1.27$ from V-1ph method\cite{Leyronas,Punk,ChevyM4,Cui2}. Near the transitions, the molecule states ($M_4(0)$ under V-2ph and $M_2(0)$ under V-1ph) can well approximate the $Q=k_F$ states, and thus the transition between $Q=0$ and $Q=k_F$ states can indeed be interpreted as the polaron-molecule transition. This sets the nature of such first-order transition between polaron and molecule.

%***********************************
\begin{figure}[t]
\centering
\includegraphics[width=0.5\textwidth]{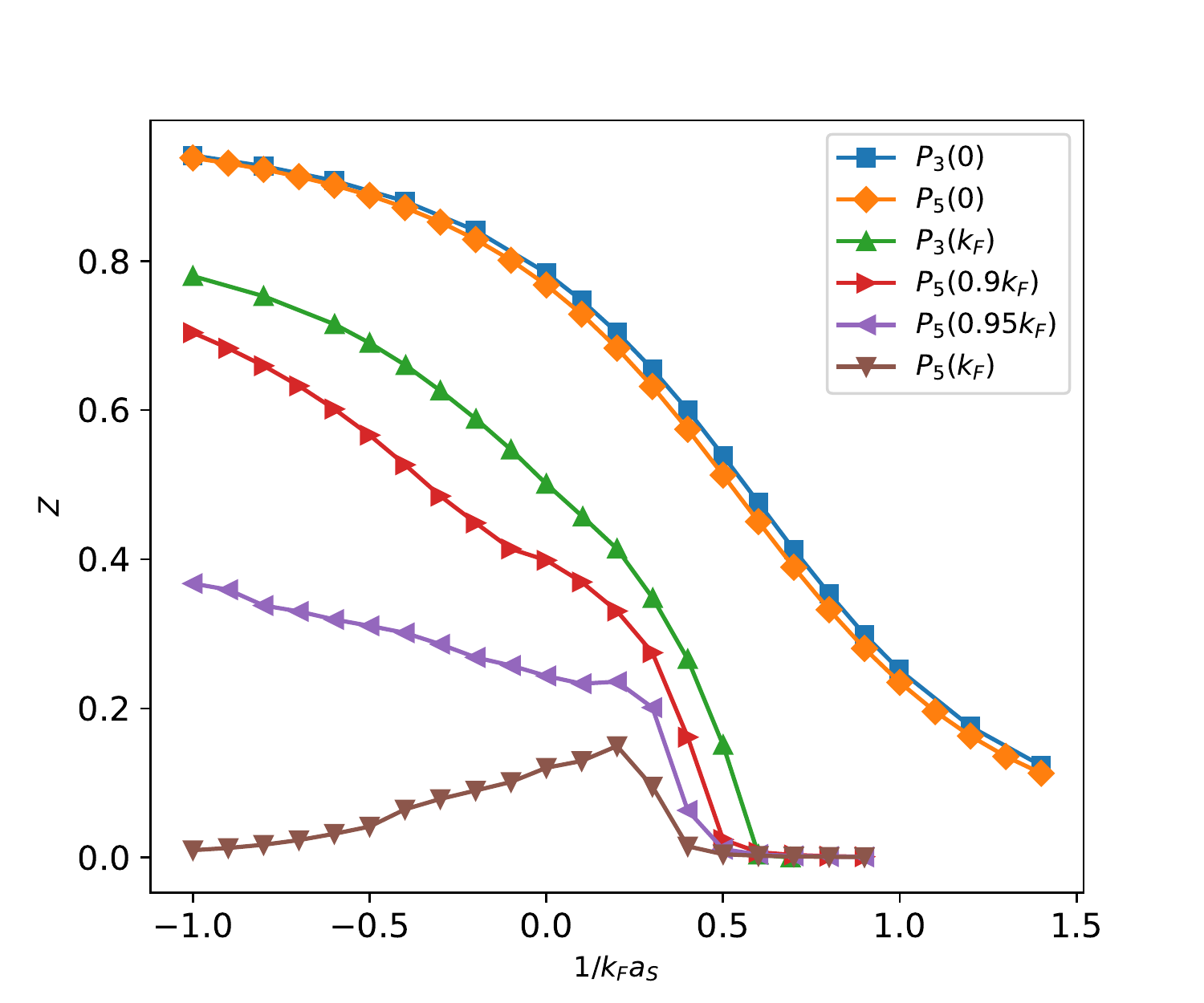}
\caption{(Color online). Residue $Z$ as a function of coupling strength $1/(k_Fa_s)$ for different momentum states using V-1ph or V-2ph methods.
}\label{fig_z}
\end{figure}
%***********************************

In Fig.\ref{fig_z}, we further show the residue $Z=|\psi_0|^2$ as a function of $1/(k_Fa_s)$ for different momentum ($Q=|{\cp Q}|$) states.   For zero-momentum $Q=0$, we can see that $Z$ is insensitive to the variational approach used (V-1ph or V-2ph). However, for momentum $Q=k_F$, $Z$ can change a lot between V-1ph and V-2ph methods, or between $P_3(k_F)$ and $P_5(k_F)$. Moreover, for a given coupling strength, $Z$ can be greatly reduced by increasing the momentum $Q$. In particular, as $Q$ approaches $k_F$, the reduction of $Z$ is quite substantial in the weak coupling limit, implying the failure of quasi-particle picture for $Q\sim k_F$ state in this regime.

In the following, we comment on the nature of polaron-molecule transition as the momentum shift by $k_F$, and its implication on the huge ground state degeneracy in the molecule limit. In Fig.\ref{fig_schematic}, we show schematically the ground state switch from polaron (${\cp Q}=0$) to molecule ($|{\cp Q}|=k_F$) as the attraction between impurity(${\downarrow}$) and majority fermions(${\uparrow}$) increases. In the extremely weak attraction limit, it is natural to expect that the ground state is a zero-momentum polaron(${\cp Q}=0$) described by a zero-momentum impurity dressed with p-h excitations in the majority Fermi sea. On the contrary, in the extremely strong attraction limit, the ground state is composed by a zero-momentum molecule on top of the rest Fermi sea. To accomplish this, the impurity has to acquire a finite momentum ${\cp  Q}$ such that it can pair with a fermion at the Fermi surface (${\cp k}_F$) to form a zero-momentum molecule (${\cp  Q}+{\cp k}_F=0$). As ${\cp k}_F$ can point to any direction on the Fermi surface, the direction of ${\cp  Q}$ is also free and the system has a huge ground state degeneracy ($SO(3)$ for 3D case) in this limit.

In fact, the huge $SO(3)$ degeneracy can also be seen clearly from the molecule dispersion  (\ref{E_M}), where the energy minimum locates at a sphere  in momentum space  with radius $|{\cp Q}|=k_F$.   Such huge degeneracy in $k$-space resembles  the single-particle SO(3) degeneracy under an isotropic spin-orbit coupling\cite{isotropic_soc_1,isotropic_soc_2}, where the ground state locates at a sphere with radius determined by the strength of spin-orbit coupling. In comparison, here the degeneracy is supported by the presence of majority Fermi sea. An important consequence of such degeneracy is that it greatly enhances the density of state (DOS) at low-energy space near $|{\cp Q}|\sim k_F$, thereby significantly favoring the molecule occupation in realistic system with a finite impurity density, as we will discuss in later section.

Given the molecule degeneracy at momentum $|{\cp Q}|=k_F$, one may raise the follow question: if  equally superpose two of the ${\cp Q}$-states, such as $|k_F{\cp e}_z\rangle+|-k_F{\cp e}_z\rangle$ that has same zero averaged momentum as the polaron state, whether there will still be the polaron-molecule transition? The answer to this question is {\it yes}. It is because that such superposed state has zero overlap with the ${\cp Q}=0$  polaron state, and thus the energy crossing between them (featuring the first-order transition) persists as changing the coupling strength. Moreover, the huge degeneracy in molecule side will not be affected since one can in principle superpose any two momentum states $|{\cp Q}_1\rangle$ and $|{\cp Q}_2\rangle$, as long as $|{\cp Q}_1|=|{\cp Q}_2|=k_F$. In fact, such superposed state is not the eigenstate of total momentum operator $\hat{P}$. Recalling that the Hamiltonian $\hat{H}$ preserves the total momentum, i.e., $[\hat{H},\hat{P}]=0$, it is a regular strategy to look for ground state as the eigen-state of both $\hat{H}$ and $\hat{P}$. In this sense, we recover the nature of polaron-molecule transition as the energy competition between different ${\cp Q}$-sectors.

\subsection{2D}

%Similar as the 3D system, we investigate the intimate relation between $M_4(0)$ and $P_5(k_F)$. We first point out that due to the logarithmic energy dependence of the scattering amplitude, it is more convenient to characterize the interaction effect using the
For 2D Fermi polaron system, we have carried out numerical simulations using both the V-2ph and V-Gph methods and found consistent results. We use the dimensionless coupling strength $\ln(k_Fa_{2d})$ to characterize the interaction effect. In our numerical calculations, we set the momentum cutoff as $k_c=30k_F$ in V-2ph method. In V-Gph method, we discretize the whole space to $40\times 40$ cells, and set the number of majority fermions as $N=49$ and the momentum cutoff as $k_c=8k_F$.

%***********************************
\begin{figure}[h]
\includegraphics[width=0.5\textwidth]{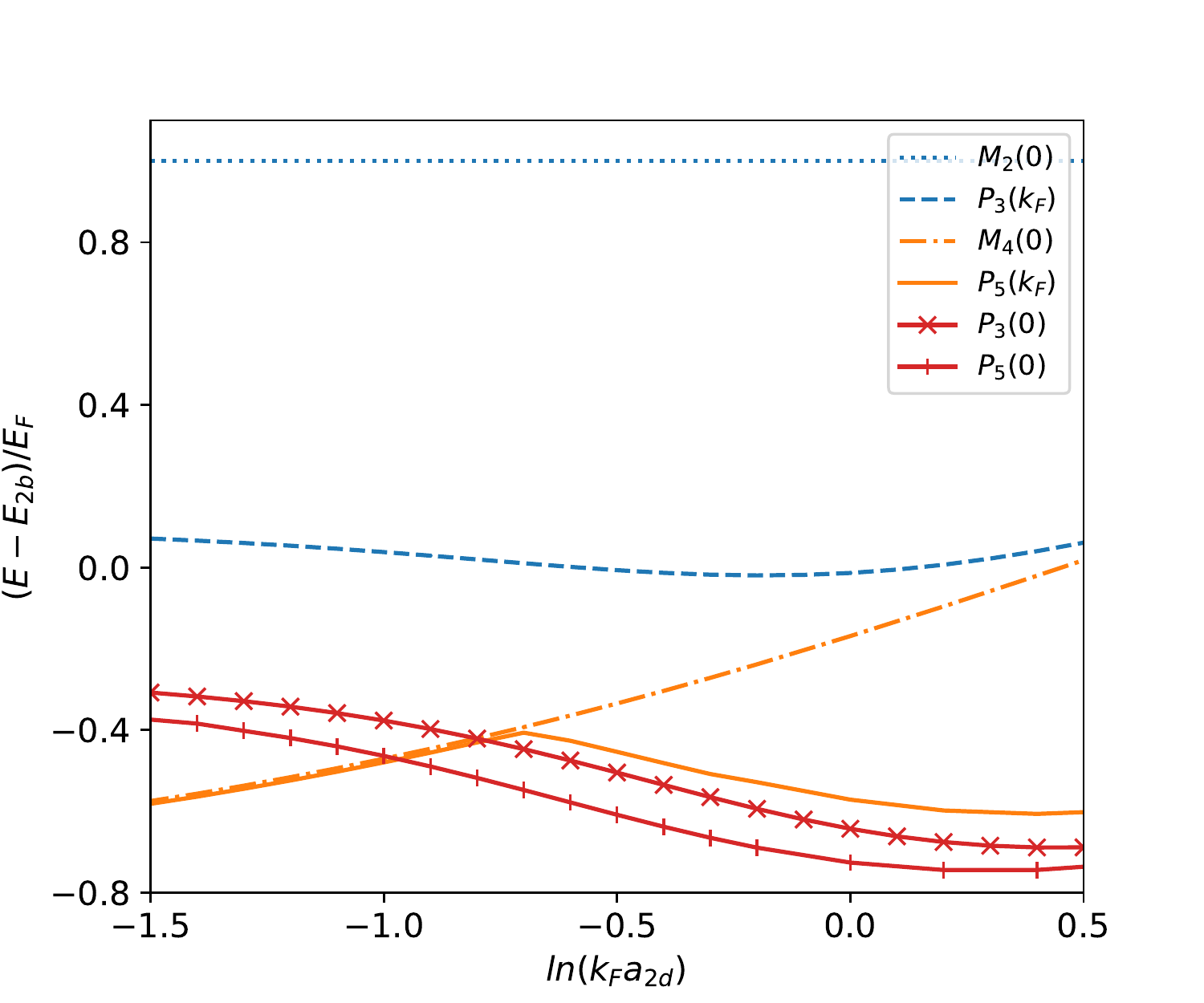}
\caption{Energy comparison in 2D. % between $P_5({\cp k}_F)$ and $M_4(0)$ from V-2ph method, and between $P_3({\cp k}_F)$ and $M_2(0)$ from V-1ph.
All energies are shifted by $E_{2b}=-1/(ma_{2d}^2)$ in order to highlight the difference.} \label{2d_1}
\end{figure}
%***********************************

\begin{figure}[h]
\includegraphics[width=0.5\textwidth]{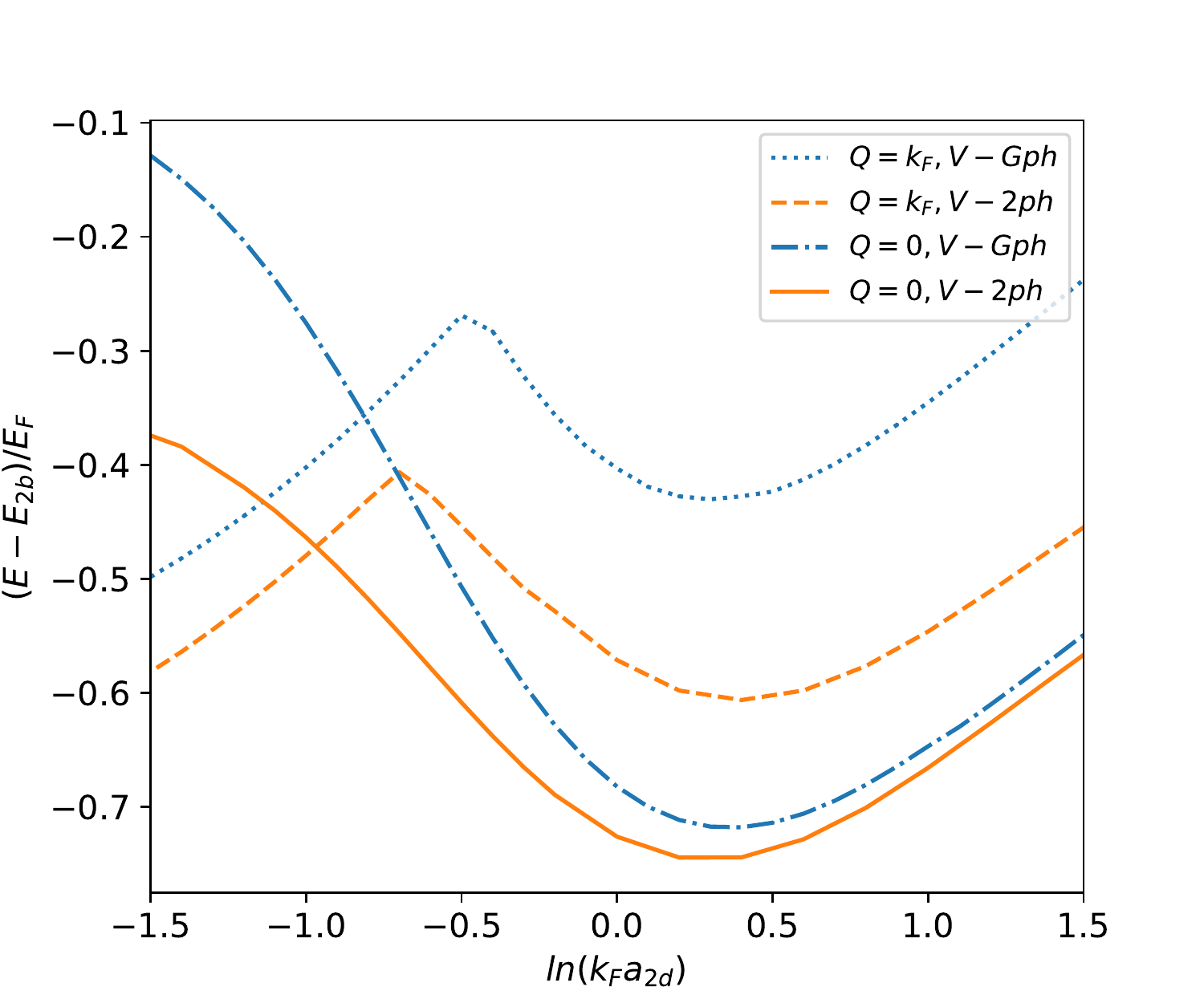}
\caption{Energies of $Q=0$ and $Q=k_F$ states as functions of coupling strengths in 2D, obtained from both the V-2ph and V-Gph methods. All energies are shifted by $E_{2b}=-1/(ma_{2d}^2)$  in order to highlight the difference.  } \label{2d_3}
\end{figure}

In Fig.\ref{2d_1},  we show the energies of  $P_5(k_F)$, $P_5(0)$ and $M_4(0)$ as functions of  $\ln(k_Fa_{2d})$,  in comparison with the energies of $P_3(k_F)$, $P_3(0)$ and $M_2(0)$. One can see that similar to the 3D case, the molecule state $M_4(0)$ always has a higher energy than $P_5({\cp k}_F)$; however, in the strong coupling regime $\ln(k_Fa_{2d})<-0.7$, the two states are indistinguishable in energy, indicating that the former can serve as a good approximation for the latter. Moreover, we note from Fig.\ref{2d_1} that the V-2ph method can produce visibly lower energy for both polaron and molecule states than V-1ph. For instance, within one p-h framework, $P_3(0)$ always has a lower energy than $P_3(k_F)$ and $M_2(0)$. However, by adding two p-h excitations, the molecule energy can be significantly reduced. In the strong coupling limit $\ln(k_Fa_{2d})\rightarrow -\infty$, the energies of $P_5(k_F)$ and $M_4(0)$ (from V-2ph) both approach $E_{2b}-E_F$, much lower than the asymptotic energy $E_{2b}+E_F$ of $P_3(k_F)$ and $M_2(0)$ states (from V-1ph) in this limit. This shows a significant role played by p-h excitations in 2D. However, adding more (three and above) p-h excitations is not expected to lower the energy too much in the strong coupling regime, since $E_{2b}-E_F$ sets the lower bound of the energy. This is further confirmed by the results from V-Gph method(see Fig.\ref{2d_3}), which includes the high-order p-h excitations and gives similar conclusion as V-2ph method, see discussions below. %Comparing to 3D case where both V-2ph and V-1ph methods produce the same asymptotic energy in strong coupling(Fig...), we conclude that the p-h excitations play a much more significant role in 2D case.

%=======================================
\begin{figure}
\includegraphics[width=0.5\textwidth]{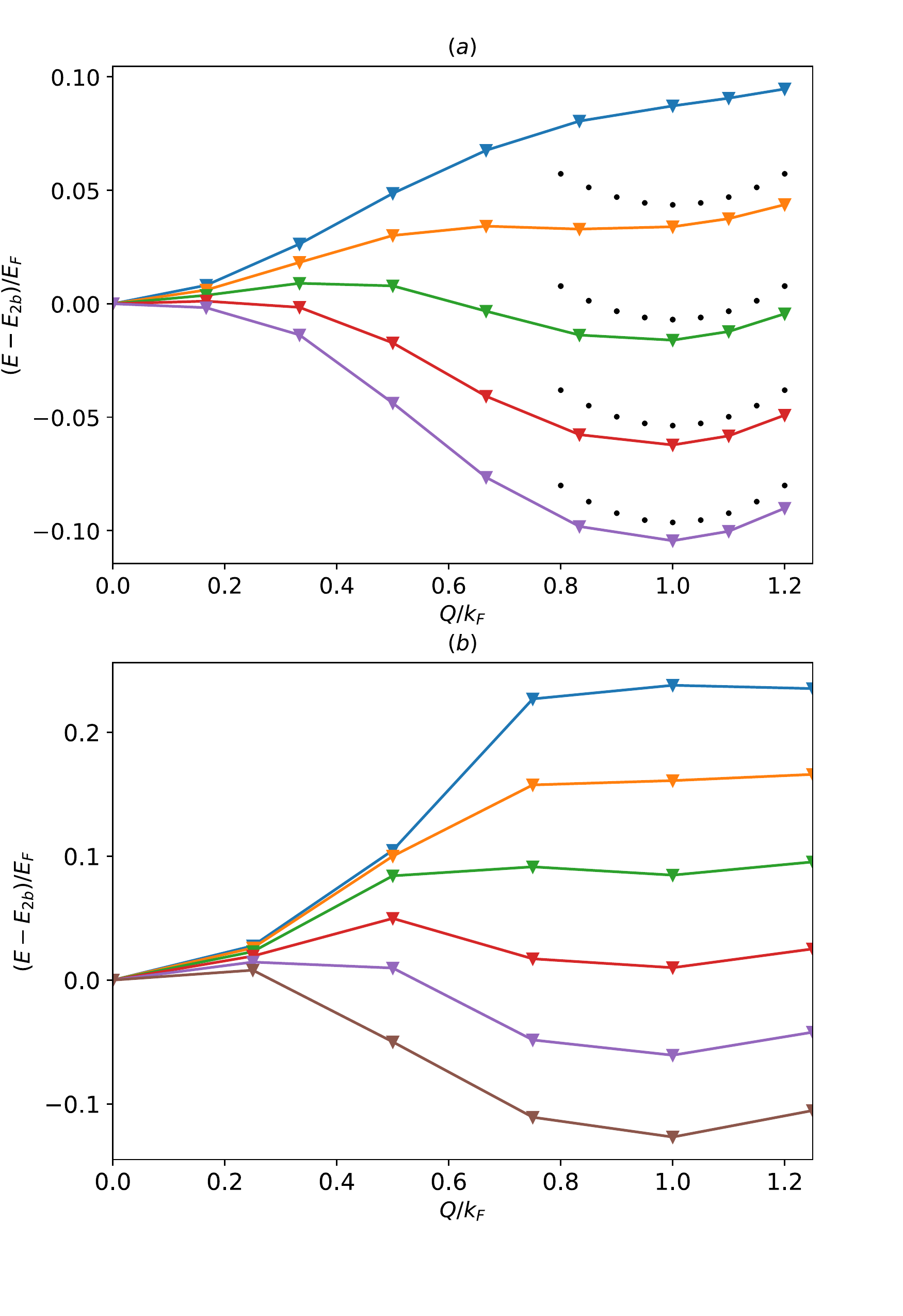}
\caption{(a)Energy dispersion of $P_5(\mathbf{Q})$(solid line) in 2D at various couplings (from top to bottom) $\ln(k_Fa_{2d})=-0.8,-0.9,-1.0,-1.1,-1.2$, shifted by the values at $\mathbf{Q}=0$.  The small black dots show the energies from $M_4({\cp Q}_{\rm M})$, with $|{\cp Q}_{\rm M}|$ shifted by $k_F$ in order to compare with the energies of $P_5(\mathbf{Q})$. (b) Energy dispersion from V-Gph method at various couplings (from top to bottom) $\ln(k_Fa_{2d})=-0.5, -0.6, -0.7, -0.8, -0.9, -1.0$, again shifted by the values at $\mathbf{Q}=0$. Here $Q=|{\cp Q}|$.} \label{2d_2}
\end{figure}
%======================================

In Fig.\ref{2d_2}(a,b), we plot out the energy dispersions at various couplings from both V-2ph and V-Gph methods, from which we see that the results from the two methods are qualitatively consistent. Namely, as increasing the attraction between impurity and fermions, there is a first-order transition at certain coupling strength where the ground state of the system switches from total momentum $Q=0$ to $Q=k_F$. Near the transition and beyond, the dispersion near $Q\sim k_F$ can indeed be well approximated by the molecule state $M_4(Q_{\rm M})$ near $Q_{\rm M}\sim 0$, see triangular points in Fig.\ref{2d_2}(a). To see more clearly the transition point, we show the energies at these two momenta as functions of coupling strengths in Fig.\ref{2d_3}.  The critical coupling at which the ground state switches from $Q=0$ to $Q=k_F$ is $\ln(k_Fa_{2d})_c\approx-0.97$ from V-2ph method, and $-0.81$ from V-Gph. In comparison, the critical coupling obtained from the comparison between $P_5(0)$ and $M_4(0)$ is $\ln(k_Fa_{2d})_c\approx-0.98$\cite{footnote,Parish2}.

%========================
%=========================

All above results confirm a first-order polaron-molecule transition in 2D single impurity system, and the nature of such transition shares the same spirit as the 3D case, i.e., the energy competition between different total momenta states ${\cp Q}=0$ and $|{\cp Q}|=k_F$. Since ${\cp Q}$ can point to any direction in the 2D plane, there will be a $SO(2)$ ground state degeneracy in the molecule regime with a fixed $|{\cp Q}|=k_F$.

We note that the polaron-molecule competition in 2D has also been investigated by Monte-Carlo methods\cite{ MC_2d_1,MC_2d_2,MC_2d_3}. Among these studies, Refs.\cite{MC_2d_1,MC_2d_2} have claimed a transition while Ref.\cite{ MC_2d_3} has claimed a smooth crossover between polaron and molecule. However, we note that in Ref.\cite{MC_2d_3} the number of majority fermions used in the weak coupling regime is different (by one) from that in the strong coupling regime. This automatically change the total momentum of the system by $k_F$ and thus the conclusion of smooth crossover is {\it not} for the same system with a fixed total momentum. Moreover, Fig.\ref{2d_3} shows that the shifted energy $E(k_F)-E_{2b}$ evolves non-monotonically with $\ln(k_Fa_{2d})$, different from the 3D case (see Fig.\ref{fig_mol_3D}). In particular, in weak coupling regime it shares similar functional lineshape as $E(0)-E_{2b}$, which may also cause the confusion that the polaron-molecule conversion in 2D is a smooth crossover.

\subsection{1D}

We will briefly go through the 1D case, where the coupling strength is governed by a dimensionless parameter $k_Fa_{1d}$, with $a_{1d}=-2/(mg)$ the 1D scattering length. In our numerical calculations, we are able to compute with different momentum cutoff $k_c$ and finally obtain the results for $k_c\rightarrow\infty$ by extrapolation.

%***********************************
\begin{figure}[h]
\includegraphics[width=0.5\textwidth]{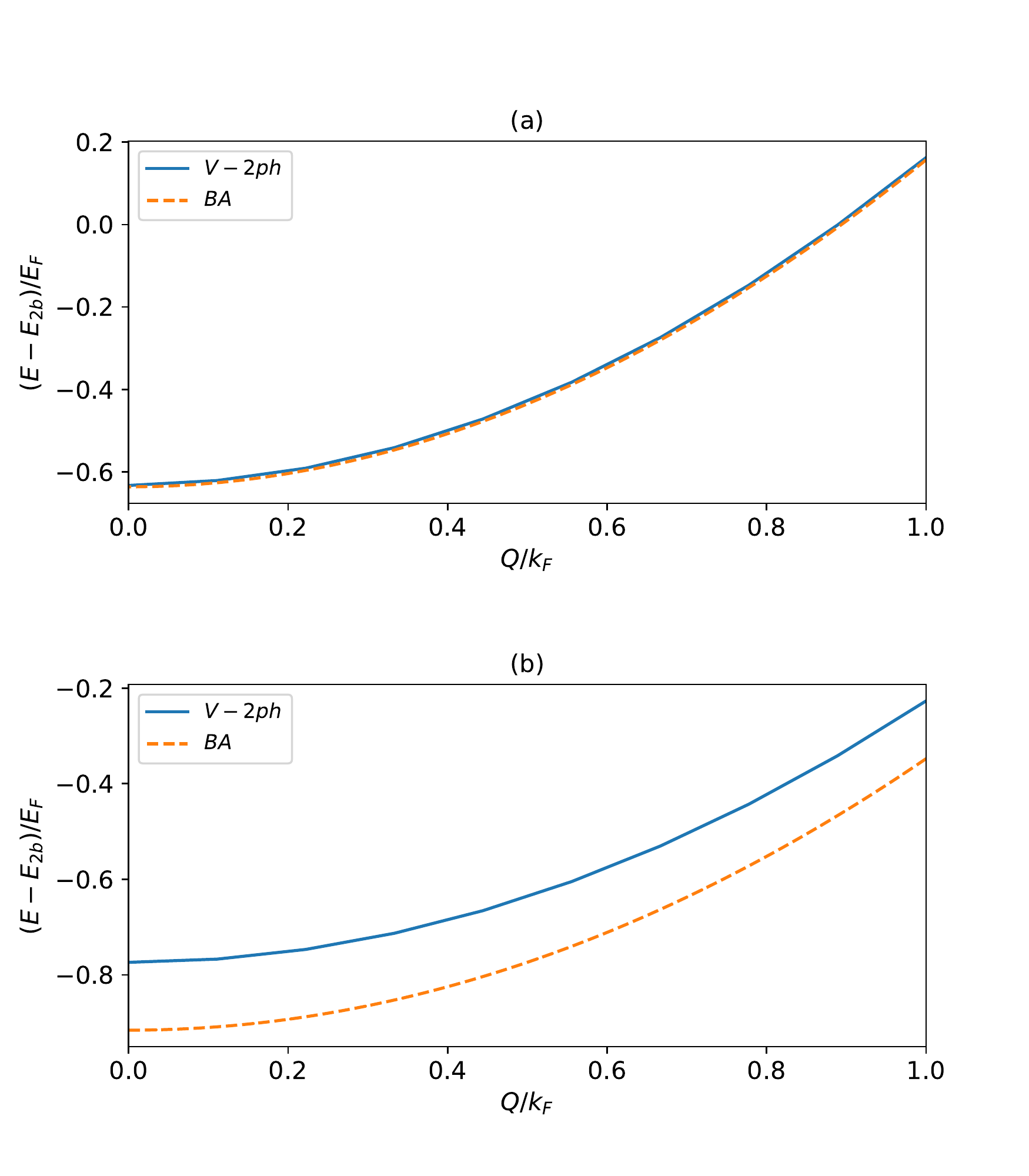}
\caption{Dispersion for 1D system at different couplings $k_Fa_{1d}=1$(a) and $0.2$(b). The solid  and dashed line are respectively from V-2ph and Bethe-ansatz\cite{McGuire,Guan,Gamayun} method, which show consistently that the ground state always stays at zero momentum. All energies are shifted by the two-body binding energy $E_{2b}=-1/(ma_{1d}^2)$. } \label{fig:1d_dis}
\end{figure}
%***********************************

In Fig.\ref{fig:1d_dis}, we show the energy dispersion at weak and strong couplings from V-2ph method (solid lines), in comparison with those from the exact Bethe ansatz solutions\cite{McGuire,Guan,Gamayun} (dashed lines). It is found that the two methods give consistent conclusion that there is no transition in the system and the ground state is always at zero momentum $Q=0$, on the contrary to 2D and 3D. Remarkably, the energy from V-2ph method fits the exact solution remarkably well in the weak coupling limit, see Fig.\ref{fig:1d_dis}(a). For strong coupling (see Fig.\ref{fig:1d_dis}(b)), the deviation between the two energies is attributed to the insufficiency of V-2ph method and thus more p-h excitations are required. In the strong coupling limit, the ground state energy(at $Q=0$) is given by $E\rightarrow E_{2b}-E_F$, signifying a smooth crossover to molecule regime for the 1D single-impurity system.

\subsection{Discussion}

In above we have shown that the presence of polaron-molecule transition sensitively depends on the dimension of the system, namely, there is such a transition in 3D and 2D but not in 1D. In the following we point out some intrinsic reasons for this sensitive dependence on dimensionality.

Let us start from the weak coupling regime that can be smoothly connected to the non-interacting limit. In this regime  one can easily anticipate  that the ground state should be the $Q=0$ polaron, describing a zero-momentum impurity dressed with a limited number of p-h excitations of background fermions. Therefore, the key question is to find out the ground state in the strong coupling regime, which determines whether there is a transition (switch of ground state) as the attraction is increased from weak to strong. Since the molecule state (belong to $Q=k_F$ sector) is an important candidate for the ground state in strong coupling regime, in the following we will analyze how its energy depends on the dimension. In particular, we will highlight the roles played by the Pauli-blocking effect and the p-h excitations of background fermions in different dimensions.

Let us consider the bare molecule $M_2(0)$ and analyze the Pauli-blocking effect to the molecule energy. For $d$-dimensional system, it has been shown that in the strong coupling or deep molecule regime (when $|E_{2b}|\rightarrow\infty$), the molecule energy (with respect to the energy of $|\textrm{FS}\rangle_N$) is\cite{Pethick}
\begin{equation}
E_M=E_{2b}-E_F+c_d E_F(2E_F/|E_{2b}|)^{(d-2)/2},
\end{equation}
with $c_d$ is a positive constant. One can see that in deep molecule regime, the shift of $E_M$ from $E_{2b}-E_F$ is negligible for 3D, a constant ($\propto E_F$) for 2D and an exceedingly large number for 1D. It means that the effect of Pauli blocking  by the underlying Fermi sea is very little for 3D molecule, but gets more and more significant if go to lower dimensions. This is because in 3D, the phase space blocked by the Fermi sea is negligible as compared to the full phase space, while in lower dimensions the difference between the two phase spaces is not that substantial. As a result, the molecule becomes energetically less favored in lower dimensions, which may serve as a crucial reason for the absence of polaron-molecule transition in 1D.

Moreover, we note that the p-h excitations also become more and more important to affect the molecule energy as going to lower-$d$ systems. As one can see from the energy comparison between $M_2(0)$ and $M_4(0)$ in Fig.\ref{fig_mol_3D} and Fig.\ref{2d_1}, adding one more p-h excitations will reduce the molecule energy by a small proportion of $E_F$ in 3D, but by a visible constant (as large as $\sim 2E_F$) in 2D. Within V-2ph, the molecule energy in 3D and 2D in strong coupling regime all approaches to $E_{2b}-E_F$ (with respect to the energy of $|\textrm{FS}\rangle_N$), which is the lowest energy one can imagine for the system. Therefore, there must be a transition between polaron ($Q=0$) and molecule ($Q=k_F$) at certain intermediate coupling strength for 3D and 2D.  On the contrary, for 1D system, the ground state is always at $Q=0$(see Fig.\ref{fig:1d_dis}), and in strong coupling regime the energy at $Q=0$ approaches $E_{2b}-E_F$ while at $Q=k_F$ approaches $E_{2b}-E_F/2$.  It means that in 1D, the polaron to molecule conversion is completed entirely within zero momentum sector, and thus the process is a smooth crossover rather than a transition.

Above analysis show that it is important to consider the effects of Pauli-blocking and p-h excitations in lower dimensional Fermi polaron systems. The interplay of these effects significantly influence the presence or absence of polaron-molecule transitions in different dimensions.

\section{Polaron-molecule coexistence and smooth crossover in realistic Fermi polaron systems}

In our previous work\cite{Cui2}, we have used the single-impurity result from V-1ph method to qualitatively explain the polaron-molecule coexistence and smooth crossover as observed in recent 3D Fermi polaron systems with a finite impurity density and at finite temperature\cite{Sagi}.
Recently, a theoretical study\cite{Parish5} extended the finite-momentum V-1ph method to finite-temperature and explained the smooth crossover between polaron and molecule. %The underlying mechanism therein is expected to be consistent with ours\cite{Cui2}.
Here we will refine the explanation by utilizing  the results from V-2ph and incorporating the trap effect through local density approximation(LDA). In our calculation, we will take the same temperature ($T=0.2T_F$) and the same impurity concentration as used in the experiment\cite{Sagi}.

As seen from  Fig.\ref{fig_transition_3D}, the double minima structure of the single-impurity dispersion  provides a clear picture of polaron-molecule coexistence under a finite  impurity density and at finite temperature. Same as Ref.\cite{Cui2}, we will neglect the thermal distortion of majority Fermi sea and mediated interactions between the impurities (including polaron-polaron, polaron-molecule and molecule-molecule interactions), which are expected to produce invisible effects at sufficiently low impurity densities. Here we only focus on two possible configurations for the dressed impurities: one is ``polaron" nearby zero-momentum and obeying fermionic statistics; the other is ``molecule" nearby $|{\cp Q}|=k_F$  and obeying bosonic statistics.

Now we discuss how to separate polaron and molecule in the dispersion curve. In the polaron-molecule coexistence regime $1/(k_Fa_s)\in(0.5,1.2)$, there is a natural momentum boundary in the diversion curve, denoted as $Q_c$, that can be chosen as the location of energy maximum between $Q=0$ and $Q=k_F$, as marked by squares  in Fig.\ref{fig_transition_3D}. After defining $Q_c$, the energy cutoff for the thermal excitation of impurities is also fixed as $E_c=E(Q_c)$. More specifically, the polaron occupies at $|{\cp Q}|<Q_c$ and the molecule occupies at $|{\cp Q}|>Q_c$ with energy cutoff $E_c$. The value of $Q_c$ outside the coexistence regime is defined as follows.  In the weak coupling regime $1/(k_Fa_s)<0.5$, the impurities occupy as polarons and there is no molecule distribution; in this case,  we define $Q_c$ as the polaron momentum when its residue reduces to $0.01$. In the strong coupling regime, $1/(k_Fa_s)>1.2$, the polaron vanishes and all impurities occupy as molecules; in this case we simply take $Q_c=0$.

Next we incorporate the trap effect. For the majority fermions, we use the zero-temperature density distribution as the approximation:
\begin{equation}
	n_{\uparrow}(\mathbf{r}) = \frac{1}{6\pi^2} \left( 2m [\mu_{\uparrow} - V(\mathbf{r})]\right)^{\frac{3}{2}},
\end{equation}
where $V(\mathbf{r})=m\omega^2{\cp r}^2/2$ is the trap potential and $\mu_{\uparrow}=k^2_{F{\uparrow}}(0)/(2m)  = (6 N_{\uparrow})^{1/3} \omega$ is chemical potential of majority fermions at the center of trap. %For the simplicity of dimensional reduction we also introduce redius of majority atoms cloud $R = \frac{2\mu}{m\omega^2}$. And
Under LDA, one can define the local Fermi momentum as $k_{F\uparrow}(\mathbf{r}) = (6\pi^2n_{\uparrow}(\mathbf{r}))^{1/3}$, which determines the local occupation of polaron and molecule states.

Under above assumptions, the local impurity density can be written as (with $\theta(x)$ step function)
	\begin{eqnarray}
		n_{\downarrow}(\mathbf{r}) &=& \int\frac{d^3{\cp Q}}{(2\pi)^3} \left[  n_F(E({\cp Q},{\cp r}),\mu_\downarrow,V_\downarrow({\cp r}))\ \theta(Q_{c}-|{\cp Q}|) \right. \nonumber\\
		&&\left.+ n_B(E({\cp Q},{\cp r}),\mu_\downarrow,V_\downarrow(r) )\ \theta(|{\cp Q}|-Q_c)\ \theta(E_c-E({\cp Q},{\cp r})) \right] \nonumber\\ \label{local_n}
	\end{eqnarray}
where $n_{F/B}(E,\mu_\downarrow,V_\downarrow(\mathbf{r})) = [1 \pm \exp(\frac{E-\mu_\downarrow+V_\downarrow(\mathbf{r})}{k_B T}) ]^{-1}$ and $V_\downarrow(\mathbf{r})= V_\uparrow(\mathbf{r}) (1- \frac{E}{E_F})$ is the renormalized trap potential felt by impurity atoms \cite{Lobo,Sagi}. Note that because of the ${\cp r}$-dependence of local $k_{F\uparrow}$, the quantities $E,\ Q_c, \ E_c$ in the above equation all depend locally on ${\cp r}$.

Following the definition of averaged density ratio in \cite{Sagi}:
\begin{equation}
\langle \frac{n_{\downarrow}}{n_{\uparrow}}\rangle=\frac{ \int d^3\mathbf{r} n_\downarrow(\mathbf{r}) \cdot \frac{n_\downarrow(\mathbf{r})}{n_\uparrow(\mathbf{r})} }{\int d^3\mathbf{r} n_\downarrow(\mathbf{r})},
\end{equation}
in our calculation we will fix $\langle \frac{n_{\downarrow}}{n_{\uparrow}}\rangle=0.15$ as in Ref.\cite{Sagi}, which is used to determine $\mu_{\downarrow}$ in Eq.~(\ref{local_n}). Then we go further to calculate trap averaged residue, contact, and the polaron energy by \begin{equation}
\begin{split}
\bar{Z} &= \frac{ \int d^3{\cp r} n_\downarrow({\cp r}) Z({\cp r}) }{\int d^3r n_\downarrow({\cp r})},\\
\bar{C} &= \frac{ \int d^3{\cp r} n_\downarrow({\cp r}) C({\cp r}) }{\int d^3r n_\downarrow({\cp r})},\\
\bar{E}_{pol} &= \frac{ \int d^3{\cp r} n_\downarrow({\cp r}) E_{pol}({\cp r}) }{\int d^3{\cp r} n_\downarrow({\cp r})},\\
\end{split}	
\end{equation}
where $Z({\cp r}),\ C({\cp r}),\ E_{pol}({\cp r})$ are :
\begin{widetext}
\begin{equation}
\begin{split}
	Z({\cp r}) &= \frac{1}{n_\downarrow({\cp r})} \int \frac{d^3 {\cp k}}{(2\pi)^3} Z({\cp k}) n_F(E({\cp k},{\cp r}),\mu_\downarrow,V_\downarrow(\mathbf{r}))\cdot\theta(Q_{c}-|{\cp k}|), \\
	C({\cp r}) &= \frac{4\pi m}{2n_\downarrow({\cp r})k_F^2}  \int \frac{d^3{\cp k}}{(2\pi)^3} \frac{dE({\cp k},{\cp r})}{d(1/k_Fa_s)} n_F(E_({\cp k},{\cp r}),\mu_\downarrow,V_\downarrow({\cp r})) \theta(Q_{c}-|{\cp k}|)\\
	\quad \quad \quad \quad &+ \frac{4\pi m}{2n_\downarrow({\cp r})k_F^2}  \int \frac{d^3{\cp k}}{(2\pi)^3} \frac{dE({\cp k},{\cp r})}{d(1/k_Fa_s)} n_B(E({\cp k},{\cp r}),\mu_\downarrow,V_\downarrow({\cp r}) )\ \theta(|{\cp k}|-Q_c)\ \theta(E_c-E({\cp k},{\cp r})),\\
	E_{pol}({\cp r}) &= E(k=0, {\cp r}).
\end{split}
\end{equation}
\end{widetext}

%==========================================
\begin{figure}[t]
\includegraphics[width=0.45\textwidth]{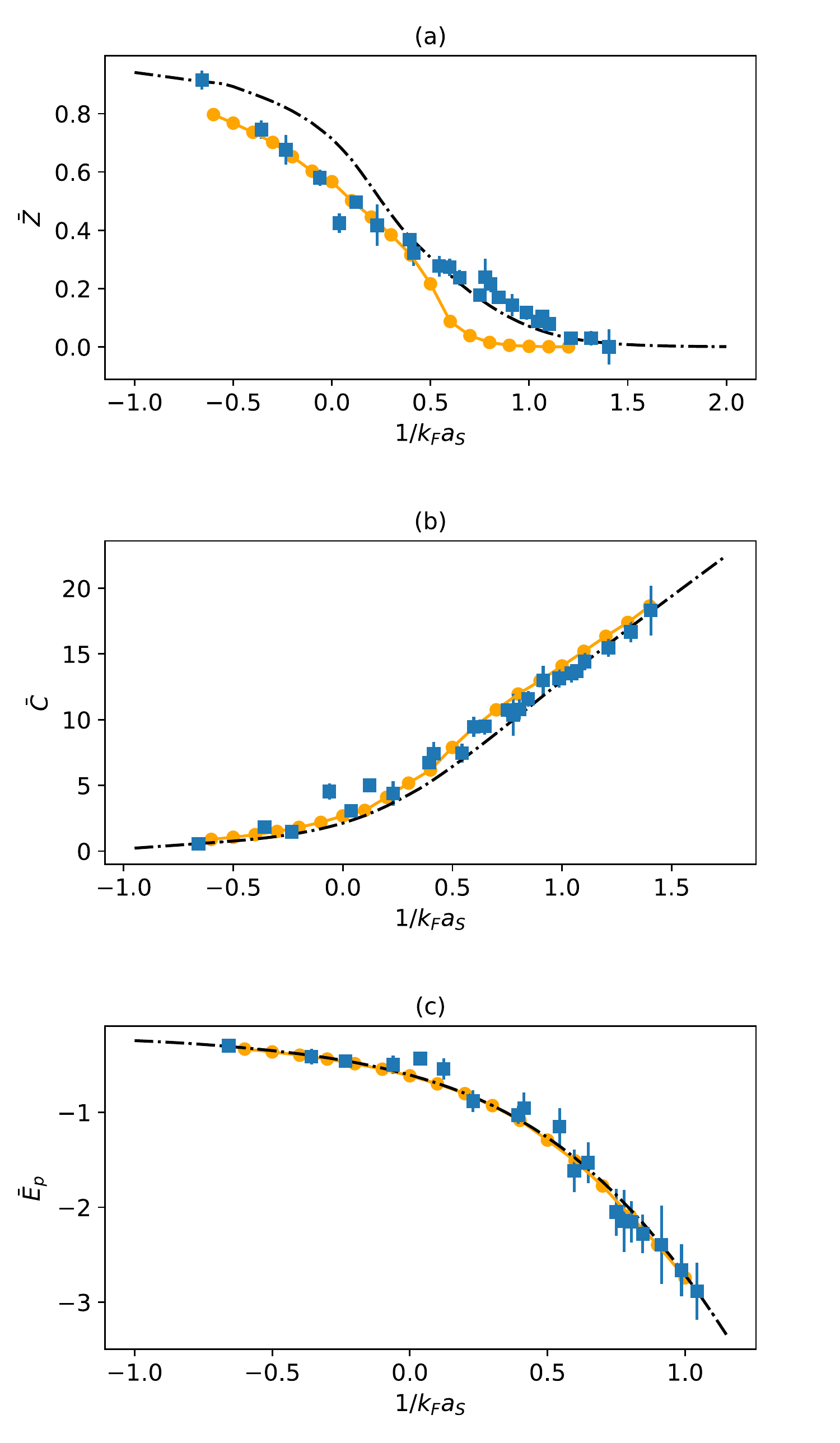}
\caption{(Color online). Residue $\bar{Z}$ (a), contact $\bar{C}$(b) and the polaron energy $\bar{E}_p$(c) as functions of coupling strength given the realistic experimental condition in Ref.\cite{Sagi}. The blue squares with error bar shows the experimental results in Ref.\cite{Sagi}. Black dashed-dot lines show the theoretical prediction in Ref.\cite{Sagi} based on the separate treatment of polaron (under V-1ph) and molecule (no p-h excitation) without considering the SO(3) degeneracy.  Orange circles and lines show our results based on V-2ph plus LDA. Here $k_F$ is the Fermi momentum of majority fermions at the trap center.  }  \label{ZEC}
\end{figure}
%=========================================

In Fig.\ref{ZEC}(a,b,c), we show the calculated $\bar{Z}$, $\bar{C}$, and $\bar{E}_p$ (see orange circles and lines) as functions of coupling strength, in comparison with the experimental data in Ref.\cite{Sagi}(shown as blue circles with error bars) and the theory prediction therein based on the separate treatment of polaron (under V-1ph) and molecule (no p-h excitation) while without considering the SO(3) degeneracy of molecules (shown by  black dashed-dot lines).
We can see that compared to the theory prediction in Ref.\cite{Sagi}, our prediction of $\bar{Z}$ is visibly lower and the prediction of $\bar{C}$ is visibly higher, giving a better fit to the experimental data in the weak coupling and near resonance regime. These visible improvements can be attributed to the following two reasons:

First, compared to the V-1ph method, the inclusion of two p-h excitations in V-2ph does not change too much the polaron energy but reduces the molecule energy considerably, see Fig.\ref{fig_mol_3D}. A direct consequence of this change is to move the polaron-molecule transition point and their coexistence region to weaker coupling side. The other consequence is to enhance the molecule occupation in the co-existence regime. These two factors both contribute to reducing the residue $\bar{Z}$ and increasing the contact $\bar{C}$ for a given coupling strength.

Secondly, we have a different classification and sampling scheme for polaron and molecule as compared to Ref.\cite{Sagi}. In particular, in the theory of Ref.\cite{Sagi} the molecule dispersion is centered at zero rather than $k_F$, and thus the SO(3) degeneracy is not considered. This significantly underestimates the molecule occupation number due to the small density of state(DoS) near ${\cp Q}\sim 0$.
%, which also causes a larger $\bar{Z}$ and a smaller $\bar{C}$ in \cite{Sagi}.
In comparison, in this work we point out that the molecule actually stays around $|{\cp Q}|\sim k_F$ with a huge $SO(3)$ degeneracy and thus a much larger DoS at low energy. This will also help to enhance the molecule occupation further and lead to a smaller $\bar{Z}$ and a larger $\bar{C}$ than the theory prediction in Ref.\cite{Sagi}. 

%==========================================
\begin{figure}[t]
\includegraphics[width=0.5\textwidth]{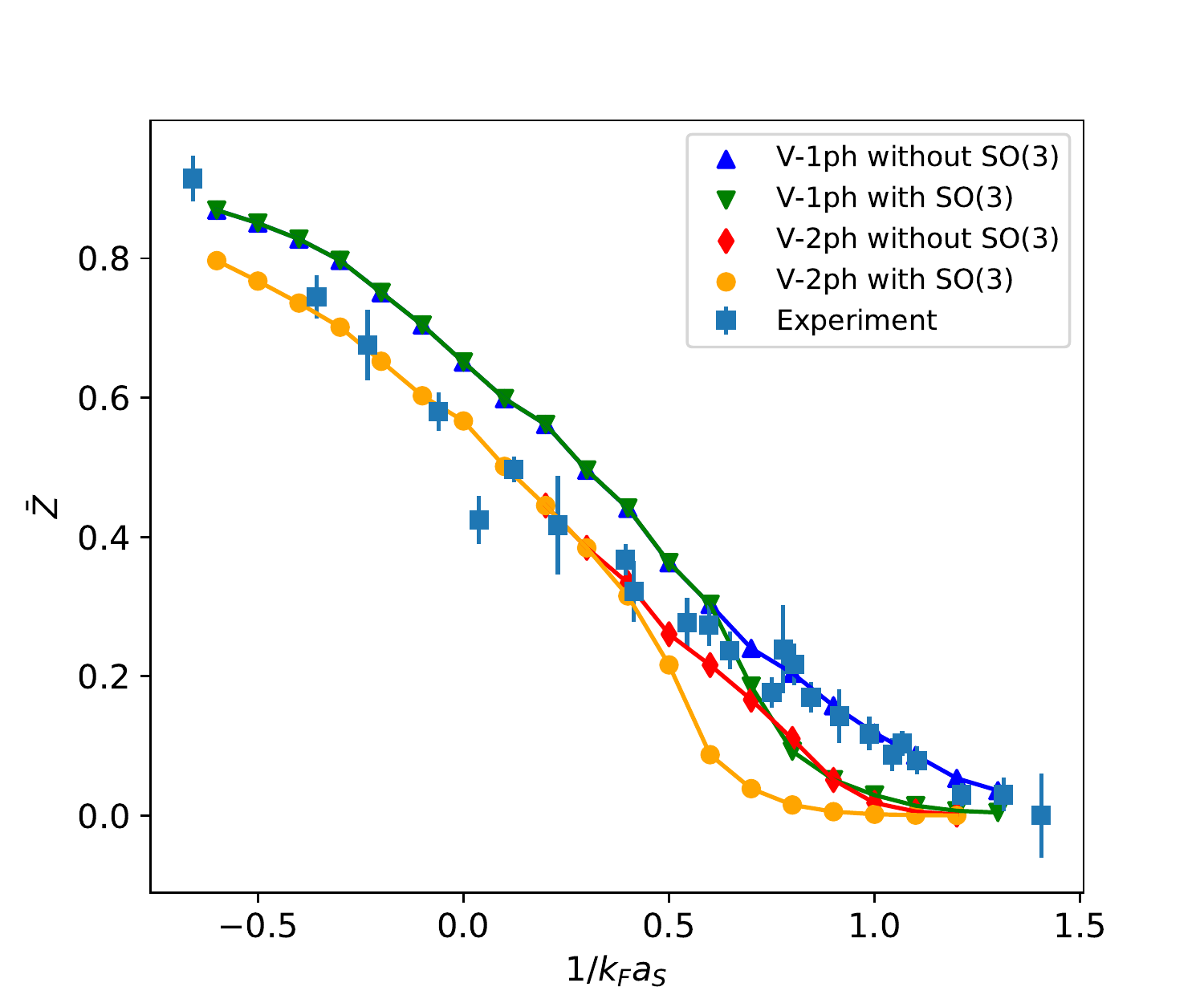}
\caption{(Color online). Residue $\bar{Z}$ from the combination of different methods (V-1ph or V-2ph) and different sampling schemes (with or without considering SO(3) degeneracy of molecules).  The blue squares with error bar shows the experimental results in Ref.\cite{Sagi}. }  \label{Z_supple}
\end{figure}
%=========================================

In order to examine the individual contribution of above two effects, in Fig.\ref{Z_supple} we plot $\bar{Z}$ as a function of coupling strength from the combination of different methods (V-1ph or V-2ph) and different sampling schemes (with or without considering SO(3) degeneracy of molecules). We can see that in the weak coupling regime, $\bar{Z}$ can be visibly reduced by   including two p-h excitations, which can be attributed to the sensitive change of $Z$ for finite-momentum polaron (note that the residue of ${\cp Q}=0$ polaron shows little difference between V-1ph and V-2ph methods, see Fig.\ref{fig_z}). In this regime, the SO(3) degeneracy takes no effect since there is no molecule occupation yet. However, when going to strong coupling regime where polaron and molecule coexist, the SO(3) degeneracy plays an important role in enhancing the molecule occupation and reducing $\bar{Z}$, regardless of the order of p-h excitations. 

Finally, it is noted that our theory does not fit well to the experimental data of $\bar{Z}$ in the strong coupling regime. For instance, $\bar{Z}$ from our prediction continuously drops to  zero around $1/(k_Fa_s)\sim 0.9$, very close to the polaron-molecule transition point ($\sim 0.91$)  for the single-impurity system. Nevertheless, the data  of $\bar{Z}$ in Ref.\cite{Sagi} show a long tail in this regime and  seem to be better fit by V-1ph results without considering the SO(3) degeneracy of molecules. Possible reasons for the discrepancy are as follows. First, the data of $\bar{Z}$ in Ref.\cite{Sagi} is not from a direct measurement; instead, it is extracted from the total Raman spectrum of impurities that is parametrized by six free parameters ($\bar{Z}$ is one of them). Moreover, 
%the parametrization of Raman signal (Eq.6 in Ref.\cite{Sagi}) assumes that the residue of polaron is independent on its momentum, which is contradictory to our finding in Fig.\ref{..}. Lastly, 
the parametrization of background Raman signal therein is based on the assumption of thermal occupation of bare molecule without SO(3) degeneracy. All of these assumptions may cause the deviation of $\bar{Z}$ from its actual value, especially in the polaron-molecule coexistence regime. Thus, the deterministic test of different theories calls for future experiment with more accurate and direct probe of various physical quantities.

%Finally, we note that our results do not fit well to the experimental data of $\bar{Z}$ near its zero crossing. As shown in Fig.\ref{ZEC}(a), $\bar{Z}$ from our prediction continuously drops to nearly zero around $1/(k_Fa_s)\sim 0.9$, very close to the polaron-molecule transition point ($\sim 0.91$)  for the single-impurity system. Nevertheless, the experimental data of $\bar{Z}$ near its zero-crossing  seem to be better fit by the V-1ph prediction. The underlying reason is unclear and awaits more exploration.

\section{Summary}

In this work we have investigated the polaron and molecule physics in 3D, 2D and 1D Fermi polaron systems by utilizing a unified variational ansatz with up to two p-h excitations(V-2ph). Moreover, we have  checked the reliability of our results by comparing with the result from the variational method in 2D based on the Gaussian sample of %selected arbitrarily
high-order p-h excitations(V-Gph), and with the result of Bethe-ansatz solutions in 1D. These methods produce consistent conclusions, which are summarized as follows:

(I) There exists a first-order transition for single-impurity system in 3D and 2D as the attraction between the impurity and fermions increases. The nature of such transition lies in an energy competition between different total momenta ${\cp Q}=0$ and $|{\cp Q}|=k_F$, with $k_F$ the Fermi momentum of majority fermions. From V-2ph method, the transition point is at $1/(k_Fa_s)=0.91$ for 3D and at $\ln(k_Fa_{2d})= -0.97$ for 2D. In 1D, there is no transition and the ground state is always at $Q=0$ for all couplings. The underlying reason for the presence/absence of such transition is analyzed to be closely related to interplay effect of Pauli-blocking and p-h excitations in different dimensions.

(II) The literally proposed molecule state has an incomplete variational space in terms of  p-h excitations, but can serve as a good approximation for the $Q=k_F$ state in strong coupling regime. Due to the finite momentum, the ground state in the molecule regime has a huge degeneracy  ($SO(3)$ for 3D and $SO(2)$ for 2D), which can greatly enhance the low-energy density of state for the molecule occupation in realistic Fermi polaron systems with a finite impurity density. Our theory well explains the coexistence and smooth crossover between polaron and molecule as observed in recent 3D Fermi polaron experiment\cite{Sagi}, and also produces quantitatively good fits to various physical quantities measured in the weak coupling and resonance regime of the system.

In the future, it would be interesting to extend our theory to various other impurity systems, such as with different mass ratios between the impurity and the background, as well as the regime with strong three-body correlations where the trimer physics can dominate.

{\it Acknowledgements.} We thank Yoav Sagi for sharing with us the data in the experiment\cite{Sagi}. This work is supported by the National Key Research and Development Program of China (2018YFA0307600, 2016YFA0300603), the National Natural Science Foundation of China (11774425, 12074419), and the Strategic Priority Research Program of Chinese Academy of Sciences (XDB33000000).


\begin{thebibliography}{99}
\bibitem{footnote} We note that Ref.\cite{Parish2} obtained the transition point between $P_5(0)$ and $M_4(0)$ as $\ln(k_Fa_{2d})_c=-0.97$. The slight difference between this result and ours is due to numerical errors.
\end{thebibliography}
\end{document}